\documentclass[twocolumn]{aastex631}

\usepackage{graphicx}
\usepackage{amsmath}
\usepackage{multirow}
\usepackage{todonotes}
\usepackage{float}

\begin{document}

\title{
Multiwavelength Investigation of $\gamma$-ray Source MGRO J1908+06 Emission Using {\it Fermi}--LAT, VERITAS, and HAWC
}

\collaboration{80}{The VERITAS collaboration}
\author[0000-0002-2028-9230]{A.~Acharyya}\affiliation{CP3-Origins, University of Southern Denmark, Campusvej 55, 5230 Odense M, Denmark}
\author[0000-0002-9021-6192]{C.~B.~Adams}\affiliation{Physics Department, Columbia University, New York, NY 10027, USA}
\author[0000-0002-3886-3739]{P.~Bangale}\affiliation{Department of Physics and Astronomy and the Bartol Research Institute, University of Delaware, Newark, DE 19716, USA}
\author[0000-0002-9675-7328]{J.~T.~Bartkoske}\affiliation{Department of Physics and Astronomy, University of Utah, Salt Lake City, UT 84112, USA}
\author[0000-0003-2098-170X]{W.~Benbow}\affiliation{Center for Astrophysics $|$ Harvard \& Smithsonian, Cambridge, MA 02138, USA}
\author[0000-0001-6391-9661]{J.~H.~Buckley}\affiliation{Department of Physics, Washington University, St. Louis, MO 63130, USA}
\author{J.~L.~Christiansen}\affiliation{Physics Department, California Polytechnic State University, San Luis Obispo, CA 94307, USA}
\author{A.~J.~Chromey}\affiliation{Center for Astrophysics $|$ Harvard \& Smithsonian, Cambridge, MA 02138, USA}
\author[0000-0003-1716-4119]{A.~Duerr}\affiliation{Department of Physics and Astronomy, University of Utah, Salt Lake City, UT 84112, USA}
\author[0000-0002-1853-863X]{M.~Errando}\affiliation{Department of Physics, Washington University, St. Louis, MO 63130, USA}
\author[0000-0002-5068-7344]{A.~Falcone}\affiliation{Department of Astronomy and Astrophysics, 525 Davey Lab, Pennsylvania State University, University Park, PA 16802, USA}
\author[0000-0001-6674-4238]{Q.~Feng}\affiliation{Department of Physics and Astronomy, University of Utah, Salt Lake City, UT 84112, USA}
\author[0000-0002-2944-6060]{G.~M.~Foote}\affiliation{Department of Physics and Astronomy and the Bartol Research Institute, University of Delaware, Newark, DE 19716, USA}
\author[0000-0002-1067-8558]{L.~Fortson}\affiliation{School of Physics and Astronomy, University of Minnesota, Minneapolis, MN 55455, USA}
\author[0000-0003-1614-1273]{A.~Furniss}\affiliation{Department of Physics, California State University - East Bay, Hayward, CA 94542, USA}
\author[0000-0002-0109-4737]{W.~Hanlon}\affiliation{Center for Astrophysics $|$ Harvard \& Smithsonian, Cambridge, MA 02138, USA}
\author[0000-0002-8513-5603]{D.~Hanna}\affiliation{Physics Department, McGill University, Montreal, QC H3A 2T8, Canada}
\author[0000-0003-3878-1677]{O.~Hervet}\affiliation{Santa Cruz Institute for Particle Physics and Department of Physics, University of California, Santa Cruz, CA 95064, USA}
\author[0000-0001-6951-2299]{C.~E.~Hinrichs}\affiliation{Center for Astrophysics $|$ Harvard \& Smithsonian, Cambridge, MA 02138, USA and Department of Physics and Astronomy, Dartmouth College, 6127 Wilder Laboratory, Hanover, NH 03755 USA}
\author[0000-0002-6833-0474]{J.~Holder}\affiliation{Department of Physics and Astronomy and the Bartol Research Institute, University of Delaware, Newark, DE 19716, USA}
\author[0000-0002-1432-7771]{T.~B.~Humensky}\affiliation{Department of Physics, University of Maryland, College Park, MD, USA and NASA GSFC, Greenbelt, MD 20771, USA}
\author[0000-0002-1089-1754]{W.~Jin}\affiliation{Department of Physics and Astronomy, University of California, Los Angeles, CA 90095, USA}
\author[0000-0002-3638-0637]{P.~Kaaret}\affiliation{Department of Physics and Astronomy, University of Iowa, Van Allen Hall, Iowa City, IA 52242, USA}
\author{M.~Kertzman}\affiliation{Department of Physics and Astronomy, DePauw University, Greencastle, IN 46135-0037, USA}
\author[0000-0003-4785-0101]{D.~Kieda}\affiliation{Department of Physics and Astronomy, University of Utah, Salt Lake City, UT 84112, USA}
\author[0000-0002-4260-9186]{T.~K.~Kleiner}\affiliation{DESY, Platanenallee 6, 15738 Zeuthen, Germany}
\author[0000-0002-4289-7106]{N.~Korzoun}\affiliation{Department of Physics and Astronomy and the Bartol Research Institute, University of Delaware, Newark, DE 19716, USA}
\author[0000-0002-5167-1221]{S.~Kumar}\affiliation{Department of Physics, University of Maryland, College Park, MD, USA }
\author[0000-0003-4641-4201]{M.~J.~Lang}\affiliation{School of Natural Sciences, University of Galway, University Road, Galway, H91 TK33, Ireland}
\author[0000-0003-3802-1619]{M.~Lundy}\affiliation{Physics Department, McGill University, Montreal, QC H3A 2T8, Canada}
\author[0000-0001-9868-4700]{G.~Maier}\affiliation{DESY, Platanenallee 6, 15738 Zeuthen, Germany}
\author{C.~E~McGrath}\affiliation{School of Physics, University College Dublin, Belfield, Dublin 4, Ireland}
\author[0000-0001-7106-8502]{M.~J.~Millard}\affiliation{Department of Physics and Astronomy, University of Iowa, Van Allen Hall, Iowa City, IA 52242, USA}
\author{J.~Millis}\affiliation{Department of Physics and Astronomy, Ball State University, Muncie, IN 47306, USA and Department of Physics, Anderson University, 1100 East 5th Street, Anderson, IN 46012}
\author[0000-0001-5937-446X]{C.~L.~Mooney}\affiliation{Department of Physics and Astronomy and the Bartol Research Institute, University of Delaware, Newark, DE 19716, USA}
\author[0000-0002-1499-2667]{P.~Moriarty}\affiliation{School of Natural Sciences, University of Galway, University Road, Galway, H91 TK33, Ireland}
\author[0000-0002-3223-0754]{R.~Mukherjee}\affiliation{Department of Physics and Astronomy, Barnard College, Columbia University, NY 10027, USA}
\author[0000-0002-6121-3443]{W.~Ning}\affiliation{Department of Physics and Astronomy, University of California, Los Angeles, CA 90095, USA}
\author[0000-0002-4837-5253]{R.~A.~Ong}\affiliation{Department of Physics and Astronomy, University of California, Los Angeles, CA 90095, USA}
\author{N.~Park}\affiliation{Department of Physics, Engineering Physics and Astronomy, Queen University, Kingston, ON K7L 3N6, Canada}
\author[0000-0001-7861-1707]{M.~Pohl}\affiliation{Institute of Physics and Astronomy, University of Potsdam, 14476 Potsdam-Golm, Germany and DESY, Platanenallee 6, 15738 Zeuthen, Germany}
\author[0000-0002-0529-1973]{E.~Pueschel}\affiliation{Fakult\"at f\"ur Physik \& Astronomie, Ruhr-Universit\"at Bochum, D-44780 Bochum, Germany}
\author[0000-0002-4855-2694]{J.~Quinn}\affiliation{School of Physics, University College Dublin, Belfield, Dublin 4, Ireland}
\author{P.~L.~Rabinowitz}\affiliation{Department of Physics, Washington University, St. Louis, MO 63130, USA}
\author[0000-0002-5351-3323]{K.~Ragan}\affiliation{Physics Department, McGill University, Montreal, QC H3A 2T8, Canada}
\author[0000-0002-7523-7366]{D.~Ribeiro}\affiliation{School of Physics and Astronomy, University of Minnesota, Minneapolis, MN 55455, USA}
\author{E.~Roache}\affiliation{Center for Astrophysics $|$ Harvard \& Smithsonian, Cambridge, MA 02138, USA}
\author[0000-0001-6662-5925]{J.~L.~Ryan}\affiliation{Department of Physics and Astronomy, University of California, Los Angeles, CA 90095, USA}
\author[0000-0003-1387-8915]{I.~Sadeh}\affiliation{DESY, Platanenallee 6, 15738 Zeuthen, Germany}
\author[0000-0002-3171-5039]{L.~Saha}\affiliation{Center for Astrophysics $|$ Harvard \& Smithsonian, Cambridge, MA 02138, USA}
\author{G.~H.~Sembroski}\affiliation{Department of Physics and Astronomy, Purdue University, West Lafayette, IN 47907, USA}
\author[0000-0002-9856-989X]{R.~Shang}\affiliation{Department of Physics and Astronomy, Barnard College, Columbia University, NY 10027, USA}
\author[0000-0003-3407-9936]{M.~Splettstoesser}\affiliation{Santa Cruz Institute for Particle Physics and Department of Physics, University of California, Santa Cruz, CA 95064, USA}
\author{A.~K.~Talluri}\affiliation{School of Physics and Astronomy, University of Minnesota, Minneapolis, MN 55455, USA}
\author{J.~V.~Tucci}\affiliation{Department of Physics, Indiana University-Purdue University Indianapolis, Indianapolis, IN 46202, USA}
\author[0000-0002-8090-6528]{J.~Valverde}\affiliation{Department of Physics, University of Maryland, Baltimore County, Baltimore MD 21250, USA and NASA GSFC, Greenbelt, MD 20771, USA}
\author{V.~V.~Vassiliev}\affiliation{Department of Physics and Astronomy, University of California, Los Angeles, CA 90095, USA}
\author{A.~Weinstein}\affiliation{Department of Physics and Astronomy, Iowa State University, Ames, IA 50011, USA}
\author[0000-0003-2740-9714]{D.~A.~Williams}\affiliation{Santa Cruz Institute for Particle Physics and Department of Physics, University of California, Santa Cruz, CA 95064, USA}
\author[0000-0002-2730-2733]{S.~L.~Wong}\affiliation{Physics Department, McGill University, Montreal, QC H3A 2T8, Canada}
\author[0009-0001-6471-1405]{J.~Woo}\affiliation{Columbia Astrophysics Laboratory, Columbia University, New York, NY 10027, USA}

\collaboration{80}{The HAWC collaboration}
\author[0000-0001-8749-1647]{R.~Alfaro}
\affiliation{Instituto de F\'isica, Universidad Nacional Autónoma de México, Ciudad de Mexico, Mexico }

\author{C.~Alvarez}
\affiliation{Universidad Autónoma de Chiapas, Tuxtla Gutiérrez, Chiapas, México}

\author{J.C.~Arteaga-Velázquez}
\affiliation{Universidad Michoacana de San Nicolás de Hidalgo, Morelia, Mexico }

\author{D.~Avila Rojas}
\affiliation{Instituto de F\'isica, Universidad Nacional Autónoma de México, Ciudad de Mexico, Mexico }

\author{R.~Babu}
\affiliation{Department of Physics, Michigan Technological University, Houghton, MI, USA }

\author[0000-0003-3207-105X]{E.~Belmont-Moreno}
\affiliation{Instituto de F\'isica, Universidad Nacional Autónoma de México, Ciudad de Mexico, Mexico }

\author{A.~Bernal}
\affiliation{Instituto de Astronom\'ia, Universidad Nacional Autónoma de México, Ciudad de Mexico, Mexico }

\author{K.S.~Caballero-Mora}
\affiliation{Universidad Autónoma de Chiapas, Tuxtla Gutiérrez, Chiapas, México}

\author[0000-0003-2158-2292]{T.~Capistrán}
\affiliation{Instituto de Astronom\'ia, Universidad Nacional Autónoma de México, Ciudad de Mexico, Mexico }

\author[0000-0002-8553-3302]{A.~Carramiñana}
\affiliation{Instituto Nacional de Astrof\'isica, Óptica y Electrónica, Puebla, Mexico }

\author[0000-0002-6144-9122]{S.~Casanova}
\affiliation{Instytut Fizyki Jadrowej im Henryka Niewodniczanskiego Polskiej Akademii Nauk, IFJ-PAN, Krakow, Poland }

\author[0000-0002-1132-871X]{J.~Cotzomi}
\affiliation{Facultad de Ciencias F\'isico Matemáticas, Benemérita Universidad Autónoma de Puebla, Puebla, Mexico }

\author[0000-0002-7747-754X]{S.~Coutiño de León}
\affiliation{Department of Physics, University of Wisconsin-Madison, Madison, WI, USA }

\author[0000-0001-9643-4134]{E.~De la Fuente}
\affiliation{Departamento de F\'isica, Centro Universitario de Ciencias Exactase Ingenierias, Universidad de Guadalajara, Guadalajara, Mexico }

\author{D.~Depaoli}
\affiliation{Max-Planck Institute for Nuclear Physics, 69117 Heidelberg, Germany}

\author{N.~Di Lalla}
\affiliation{Department of Physics, Stanford University: Stanford, CA 94305–4060, USA}

\author{R.~Diaz Hernandez}
\affiliation{Instituto Nacional de Astrof\'isica, Óptica y Electrónica, Puebla, Mexico }

\author[0000-0002-2987-9691]{M.A.~DuVernois}
\affiliation{Department of Physics, University of Wisconsin-Madison, Madison, WI, USA }

\author[0000-0001-7074-1726]{C.~Espinoza}
\affiliation{Instituto de F\'isica, Universidad Nacional Autónoma de México, Ciudad de Mexico, Mexico }

\author{K.L.~Fan}
\affiliation{Department of Physics, University of Maryland, College Park, MD, USA }

\author[0000-0002-5387-8138]{K.~Fang}
\affiliation{Department of Physics, University of Wisconsin-Madison, Madison, WI, USA }

\author[0000-0002-0173-6453]{N.~Fraija}
\affiliation{Instituto de Astronom\'ia, Universidad Nacional Autónoma de México, Ciudad de Mexico, Mexico }

\author[0000-0002-4188-5584]{J.A.~García-González}
\affiliation{Tecnologico de Monterrey, Escuela de Ingenier\'{i}a y Ciencias, Ave. Eugenio Garza Sada 2501, Monterrey, N.L., Mexico, 64849}

\author[0000-0003-1122-4168]{F.~Garfias}
\affiliation{Instituto de Astronom\'ia, Universidad Nacional Autónoma de México, Ciudad de Mexico, Mexico }

\author[0000-0002-5209-5641]{M.M.~González}
\affiliation{Instituto de Astronom\'ia, Universidad Nacional Autónoma de México, Ciudad de Mexico, Mexico }

\author[0000-0002-9790-1299]{J.A.~Goodman}
\affiliation{Department of Physics, University of Maryland, College Park, MD, USA }

\author{S.~Groetsch}
\affiliation{Department of Physics, Michigan Technological University, Houghton, MI, USA }

\author{S.~Hernández-Cadena}
\affiliation{Instituto de F\'isica, Universidad Nacional Autónoma de México, Ciudad de Mexico, Mexico }

\author{J.~Hinton}
\affiliation{Max-Planck Institute for Nuclear Physics, 69117 Heidelberg, Germany}

\author[0000-0002-3808-4639]{D.~Huang}
\affiliation{Department of Physics, University of Maryland, College Park, MD, USA }

\author[0000-0002-5527-7141]{F.~Hueyotl-Zahuantitla}
\affiliation{Universidad Autónoma de Chiapas, Tuxtla Gutiérrez, Chiapas, México}

\author[0000-0001-5811-5167]{A.~Iriarte}
\affiliation{Instituto de Astronom\'ia, Universidad Nacional Autónoma de México, Ciudad de Mexico, Mexico }

\author{S.~Kaufmann}
\affiliation{Universidad Politecnica de Pachuca, Pachuca, Hgo, Mexico }

\author{D.~Kieda}
\affiliation{Department of Physics and Astronomy, University of Utah, Salt Lake City, UT, USA }

\author[0000-0002-2153-1519]{J.~Lee}
\affiliation{University of Seoul, Seoul, Rep.of Korea }

\author[0000-0001-5516-4975]{H.~León Vargas}
\affiliation{Instituto de F\'isica, Universidad Nacional Autónoma de México, Ciudad de Mexico, Mexico }

\author[0000-0001-8825-3624]{A.L.~Longinotti}
\affiliation{Instituto de Astronom\'ia, Universidad Nacional Autónoma de México, Ciudad de Mexico, Mexico }

\author[0000-0003-2810-4867]{G.~Luis-Raya}
\affiliation{Universidad Politecnica de Pachuca, Pachuca, Hgo, Mexico }

\author[0000-0001-8088-400X]{K.~Malone}
\affiliation{Physics Division, Los Alamos National Laboratory, Los Alamos, NM, USA }

\author[0000-0002-2824-3544]{J.~Martínez-Castro}
\affiliation{Centro de Investigaci'on en Computaci'on, Instituto Polit'ecnico Nacional, M'exico City, M'exico.}

\author[0000-0002-2610-863X]{J.A.~Matthews}
\affiliation{Dept of Physics and Astronomy, University of New Mexico, Albuquerque, NM, USA }

\author{P.~Miranda-Romagnoli}
\affiliation{Universidad Autónoma del Estado de Hidalgo, Pachuca, Mexico }

\author[0000-0001-9361-0147]{J.A.~Morales-Soto}
\affiliation{Universidad Michoacana de San Nicolás de Hidalgo, Morelia, Mexico }

\author[0000-0002-1114-2640]{E.~Moreno}
\affiliation{Facultad de Ciencias F\'isico Matemáticas, Benemérita Universidad Autónoma de Puebla, Puebla, Mexico }

\author[0000-0002-7675-4656]{M.~Mostafá}
\affiliation{Department of Physics, Temple University, Philadelphia, PA, USA }

\author[0000-0003-1059-8731]{L.~Nellen}
\affiliation{Instituto de Ciencias Nucleares, Universidad Nacional Autónoma de Mexico, Ciudad de Mexico, Mexico }

\author[0000-0001-5998-4938]{E.G.~Pérez-Pérez}
\affiliation{Universidad Politecnica de Pachuca, Pachuca, Hgo, Mexico }

\author[0000-0002-6524-9769]{C.D.~Rho}
\affiliation{University of Seoul, Seoul, Rep.of Korea }

\author[0000-0003-1327-0838]{D.~Rosa-González}
\affiliation{Instituto Nacional de Astrof\'isica, Óptica y Electrónica, Puebla, Mexico }

\author{H.~Salazar}
\affiliation{Facultad de Ciencias F\'isico Matemáticas, Benemérita Universidad Autónoma de Puebla, Puebla, Mexico }

\author{A.~Sandoval}
\affiliation{Instituto de F\'isica, Universidad Nacional Autónoma de México, Ciudad de Mexico, Mexico }

\author[0000-0001-8644-4734]{M.~Schneider}
\affiliation{Department of Physics, University of Maryland, College Park, MD, USA }

\author{J.~Serna-Franco}
\affiliation{Instituto de F\'isica, Universidad Nacional Autónoma de México, Ciudad de Mexico, Mexico }

\author{Y.~Son}
\affiliation{University of Seoul, Seoul, Rep.of Korea }

\author[0000-0002-1492-0380]{R.W.~Springer}
\affiliation{Department of Physics and Astronomy, University of Utah, Salt Lake City, UT, USA }

\author{O.~Tibolla}
\affiliation{Universidad Politecnica de Pachuca, Pachuca, Hgo, Mexico }

\author[0000-0001-9725-1479]{K.~Tollefson}
\affiliation{Department of Physics and Astronomy, Michigan State University, East Lansing, MI, USA }

\author[0000-0002-1689-3945]{I.~Torres}
\affiliation{Instituto Nacional de Astrof\'isica, Óptica y Electrónica, Puebla, Mexico }

\author{R.~Torres-Escobedo}
\affiliation{Tsung-Dao Lee Institute, Shanghai Jiao Tong University, Shanghai, China}

\author{R.~Turner}
\affiliation{Department of Physics, Michigan Technological University, Houghton, MI, USA }

\author{F.~Ureña-Mena}
\affiliation{Instituto Nacional de Astrof\'isica, Óptica y Electrónica, Puebla, Mexico }

\author{E.~Varela}
\affiliation{Facultad de Ciencias F\'isico Matemáticas, Benemérita Universidad Autónoma de Puebla, Puebla, Mexico }

\author{X.~Wang}
\affiliation{Department of Physics, Michigan Technological University, Houghton, MI, USA }

\author[0000-0003-0513-3841]{H.~Zhou}
\affiliation{Tsung-Dao Lee Institute, Shanghai Jiao Tong University, Shanghai, China}

\collaboration{80}{The {\it Fermi}--LAT collaboration}
\author[0000-0001-9633-3165]{J.~Eagle}
\affiliation{NASA Goddard Space Flight Center, 
Greenbelt, MD 20781, USA}
\author{S.~Kumar}
\affiliation{Department of Physics and Astronomy, University of Maryland, 
College Park, MD 20742, USA}

\correspondingauthor{Ruo-Yu Shang}
\email{r.y.shang@gmail.com}
\correspondingauthor{Jordan Eagle}
\email{jordan.l.eagle@nasa.gov}
\correspondingauthor{Sajan Kumar}
\email{sajkumar@udel.edu}
\correspondingauthor{S.~Coutiño De León}
\email{scoutino@icecube.wisc.edu}

\begin{abstract}

This paper investigates the origin of the $\gamma$-ray emission from MGRO J1908+06 in the GeV-TeV energy band.
By analyzing the data collected by {\it Fermi}--LAT, VERITAS, and HAWC, with the addition of spectral data previously reported by LHAASO \citep{cao2021ultrahigh,cao2023first}, a multiwavelength (MW) study of the morphological and spectral features of MGRO J1908+06 provides insight into the origin of the $\gamma$-ray emission. 
The mechanism behind the bright TeV emission is studied by constraining the magnetic field strength, the source age and the distance through detailed broadband modeling.
Both spectral shape and energy-dependent morphology support the scenario that inverse-Compton (IC) emission of an evolved pulsar wind nebula (PWN) associated with PSR J1907+0602 is responsible for the MGRO J1908+06 $\gamma$-ray emission with a best-fit true age of $T=22\pm 9$ kyr and a magnetic field of $B=5.4 \pm 0.8\ \mu\mathrm{G}$, assuming the distance to the pulsar $d_{\mathrm{PSR}}=3.2$ kpc.

\end{abstract}

\keywords{MGRO J1908+06 --- PSR J1907+0602 --- pulsar wind nebulae --- $\gamma$-ray astronomy}

\section{Introduction}
\label{sec:intro}

In recent years, the HAWC and the LHAASO collaborations have reported the detection of energetic Galactic sources that are capable of emitting photons with energies above dozens of TeV, and many of these are spatially associated with pulsar wind nebulae \citep{abeysekara2020multiple,cao2021ultrahigh,cao2023first}.
The TeV detections of several PWNe by HAWC and LHAASO confirm that these objects are effective in accelerating particles up to the PeV regime.
The acceleration mechanism and particle transport within PWNe can be understood through spectral and morphological studies. 
The age of the pulsar, the magnetic field inside the PWN, and the material surrounding the system such as the host supernova remnant (SNR) shell, are all key parameters for understanding the underlying physical processes and particle acceleration.
The nearby middle-aged (10-100 kyr) PWNe are particularly interesting because these objects are close enough that GeV-TeV instruments can resolve complex morphologies, thus revealing the interaction of the PWN with the host SNR, which is expected to take place for the middle-aged systems, and particle radiative and transport processes.

\begin{figure}
\centering
\includegraphics[width=\linewidth]
{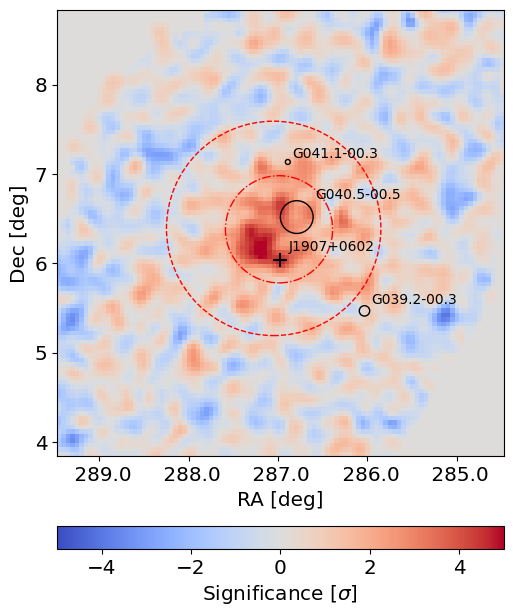}
\caption{
VERITAS $\gamma$-ray significance map in the energy range $[0.5,7.9]$ TeV.
The black cross indicates the location of PSR J1907+0602, and the black circles outline the shell sizes of the SNRs in the region, including (middle) G40.5-0.5, (top) G041.1-00.3, and (bottom) G039.2-00.3.
The red dashed circle shows the region of interest in which the VERITAS spectrum is extracted from, and the red dashdot circle shows the best-fit Gaussian radius in the Fermi--LAT data. The color scale is limited to a range of $(-5,+5)$.
}
\label{fig:intro_map}
\end{figure}

MGRO J1908+06 was first discovered by the Milagro experiment \citep{abdo2007tev} at a median energy of $\sim 20$ TeV.
The source features bright high-energy emission of about $1.36\ \text{C.U.}$ (Crab Nebula unit) at 100 TeV \citep{cao2021ultrahigh} with extension reported by multiple TeV instruments including H.E.S.S. \citep[$0.34^{\circ}$ in radius,][]{aharonian2009detection}, VERITAS \citep[$0.44^{\circ}$ in radius,][]{aliu2014investigating}, ARGO-YBJ6 \citep[$0.49^{\circ}$ in radius,][]{bartoli2012observation}, HAWC \citep[$0.52^{\circ}$ in radius,][]{abeysekara2020multiple}, and LHAASO \citep[$0.43^{\circ}$ in radius,][]{cao2021ultrahigh}.
A recent report by \cite{kostunin2021revisiting} further showed that the source displays an energy-dependent morphology.
While the origin of the MGRO J1908+06 emission remains unknown, the supernova remnant SNR~G40.5-0.5 and pulsar (PSR J1907+0602) have been suggested by recent literature as the potential counterparts to the TeV emission \citep[see e.g.,][]{aliu2014investigating}.
SNR G40.5-0.5, located at R.A., Dec. $= 286.786^{\circ}, 6.498^{\circ}$ (J2000), has an estimated age of 20-40 kyr and an estimated distance of 5.5-8.5 kpc based on the relation of the surface brightness and the diameter of the SNR \citep{downes1980g40}.
A recent discovery of a radio pulsar PSR J1907+0631 located near the center of the SNR suggests a distance of 7.9 kpc based on the dispersion measure of the pulsar \citep{lyne2017timing}. 
Different distance estimations have also been suggested by associating molecular clouds with the remnant, including a distance of 3.4 kpc by \cite{yang2006molecular} and a distance of 8.7 kpc by \cite{duvidovich2020radio}.
The nearby pulsar PSR J1907+0602, located at R.A., Dec. $= 286.978^{\circ}, 6.038^{\circ}$ (J2000), was originally detected by the {\it Fermi} Large Area Telescope ({\it Fermi}--LAT) and has a spin-down power of $\dot{E}=2.8\times 10^{36}\ \text{erg/s}$, a characteristic age of $t_{\mathrm{c}} = 19.5$ kyr, and an estimated distance of $3.2 \pm 0.6$ kpc \citep{abdo2010psr}.
The pulsar, PSR J1907+0602, has been observed by the Chandra X-ray Observatory \citep{abdo2010psr} and the XMM-Newton X-ray satellite \citep{pandel2012multi, li2021investigating}.
\cite{pandel2012multi} analyzed a 52-ks XMM-Newton observation and found a marginal excess separated from the pulsar by a distance of $7''$, which could be interpreted as a bow shock in front of the pulsar as it moved to its current location from the centroid of MGRO J1908+06.
\cite{abdo2010psr} reported some evidence for the spatial extent of the X-ray emission based on a 19-ks Chandra observation, which suggested a compact pulsar wind nebula.
However, using a deeper exposure of 109-ks of XMM-Newton data, \cite{li2021investigating} found no extended nebula emission and that the X-ray emission is consistent with a point-like pulsar.
The locations of PSR J1907+0602 and of SNR G40.5-0.5 can be found in Figure \ref{fig:intro_map} overlaid with the TeV emission morphology.

On the origin of MGRO J1908+06 emission, a recent study of radio data by \cite{duvidovich2020radio} found molecular clouds around the border of SNR G40.5-0.5 and suggested that the TeV emission is a combination of a leptonic component emitted by the PWN of PSR J1907+0602 and a hadronic component produced by the interaction between the molecular clouds and SNR G40.5-0.5.
The two-component hypothesis is further explored in a multiwavelength analysis by \cite{crestan2021multiwavelength}, in which the authors concluded that the analysis result favors the two-component scenario with the GeV emission (10-1000 GeV) being leptonic, and the TeV emission having a hadronic origin.
The two-component scenario is also supported by a spectral analysis including HAWC and LHAASO data \citep{de2022exploring}.
In addition to the TeV emission, extended GeV emission at the vicinity of PSR J1907+0602 is found in an analysis of {\it Fermi}--LAT data by \citet{li2021investigating}, where a low-energy ($<10\ \text{GeV}$) component is said to be hadronic emission originating from the interaction between the SNR and nearby molecular clouds, while the high-energy ($>10\ \text{GeV}$) component is likely to be the inverse Compton emission of the PWN associated to PSR J1907+0602.

While it has been indicated by previous works that both the PWN and the nearby SNR are plausible contributors to both the GeV and TeV emission, it remains unknown to what degree.
This paper reports a new energy-dependent morphological study of MGRO J1908+06 using the {\it Fermi}--LAT, VERITAS, and HAWC observations and multiwavelength spectral modeling across the GeV-TeV energy domain to provide new insights into the origin of the MGRO J1908+06 emission, taking into consideration as well the absence of an X-ray detection of the nebula.
Section \ref{sec:fermi_analysis}, \ref{sec:veritas_analysis}, \ref{sec:hawc_analysis} describe the {\it Fermi}--LAT, VERITAS, and HAWC data analyses and results, respectively.
Section \ref{sec:mgro_j1908_models} explores possible scenarios for the MGRO J1908+06 emission, including a leptonic explanation in Section \ref{sec:leptonic_model} and a hadronic explanation in Section \ref{sec:hadronic_model}. 
The preferred scenarios for the MGRO J1908+06 are summarized in Section \ref{sec:discussion}.

\section{{\it Fermi}--LAT data analysis}\label{sec:fermi_analysis}

The principal scientific instrument on the Fermi Gamma-ray Space Telescope is the Large Area Telescope \citep[LAT,][]{atwood2009}. The LAT instrument is sensitive to $\gamma$-rays with energies from 50\,MeV to $> 300$\,GeV \citep{4fgl2020}. The instantaneous field of view of LAT is $\sim 2.4$ steradian and has been continuously surveying the entire sky every 3\,hours since beginning operation in 2008 August. The on-axis effective area at $1\ \mathrm{GeV}$ is close to $8000\ \mathrm{cm^{2}}$, rises to $9000\ \mathrm{cm^{2}}$ at $2\ \mathrm{GeV}$ and remains constant until $500 \ \mathrm{GeV}$. The point spread function (PSF, $68\%$ containment radius) at $1\,\mathrm{GeV}$ is $< 1^{\circ}$ and becomes $\sim 0.1^{\circ}$ above $10 \ \mathrm{GeV}$.

We analyze just over 14\, years (from 2008 August to 2022 November) of Pass~8 \texttt{SOURCE} class data \citep{atwood2013,pass82018} between 30\,GeV and 2\,TeV, similar to \cite{li2021investigating}, but with 3 more years of {\it Fermi}--LAT data. The energy range for the analysis is motivated by avoiding contamination from the $\gamma$-ray pulsar J1907+0602 below 30\,GeV in addition to the possible hadronic component $E<10\,$GeV as presented in \citet{li2021investigating}. Photons detected at zenith angles larger than 90\,$^\circ$ were excluded to limit the contamination from $\gamma$-rays generated by cosmic ray (CR) interactions in the upper layers of the Earth's atmosphere. We do not require pulsar gating of the GeV PSR~J1907+0602, since the magnetospheric emission from PSR~J1907+0602 exhibits a spectral cutoff at 2.9 GeV and is not significant at energies above 30\,GeV \citep{abdo2013second,li2021investigating}. Aside from the Fermi--LAT detection of the pulsar, there is no Fermi counterpart in the 4FGL catalog associated with MGRO~J1908+06 \citep{4fgldr3}.

We perform a joint likelihood analysis of events according to their PSF type (\texttt{PSF0, PSF1, PSF2}, and \texttt{PSF3}) using the latest Fermitools package\footnote{\url{https://fermi.gsfc.nasa.gov/ssc/data/analysis/software/}} (v.2.2.11) and FermiPy Python~3 package \citep[v.1.2.0][]{fermipy2017}. Details of the analysis are provided in Appendix~\ref{sec:fermi_details}.

Figure~\ref{fig:vts_flux_map} (a) displays the 30 -- 300\,GeV residual excess counts map centered on MGRO~J1908+06. The residual excess counts map shows significant unmodeled residuals coincident to the MGRO~J1908+06 source. 
The excess corresponds to a 3\,$\sigma$ significance test statistic (TS, see Appendix~\ref{sec:fermi_details}).

\subsection{{\it Fermi}--LAT Data Analysis Results}
To model the $\gamma$-ray emission excess coincident with MGRO~J1908+06, we first add a point source at the TeV PWN location R.A., Dec. = 287.11\,$^\circ$, +6.21\,$^\circ$ (J2000) to the 30\,GeV--2\,TeV source model. We set the spectrum to a power law,
\begin{equation}
  \frac{dN}{dE} = N_{0} \left(\frac{E}{E_0}\right)^{-\Gamma_{\gamma}},
\end{equation}
where $E_0$ is set to 1000\,MeV. We allow the spectral index and normalization to vary. We localize the point source with \texttt{GTAnalysis.localize} to find the best-fit position and uncertainty. The localized position for the new $\gamma$-ray source is offset by 0.4\,$^\circ$ from the exact position of PSR~J1907+0602, corresponding to the nearby peak excess (Fig~\ref{fig:vts_flux_map} (a)) and has R.A., Dec. = 287.25\,$^\circ$, +6.21\,$^\circ$ (J2000). The corresponding 95\% positional uncertainty radius is {\bf $r_{95}=0.10 ^\circ$}. The TS of the $\gamma$-ray source is 16 at this location with a spectral index $\Gamma_\gamma = 2.50 \pm 0.40$, corresponding to a significance at the 3\,$\sigma$ level. 

\begingroup
\begin{table*}[!htb]
\centering
\begin{tabular}{cccccc}
\hline
\hline
\ Spatial Template & TS & TS$_{\text{ext}}$ & (R.A., Dec.) ($^\circ$, J2000) & $r$ or $\sigma$ ($^\circ$) & 95\% U.L. ($^\circ$) \
\\
\hline
Point Source & 15.8 & -- & 287.25, +6.21 & -- & -- \\
Radial Disk & 38.4 & 21.9 & 286.87, +6.30 & 0.44 $\pm$ 0.03 & 0.48 \\
Radial Gaussian & 37.4 & 34.9 & 286.99, +6.38 & 0.60 $\pm$ 0.10 & 0.81 \\
\hline
\hline
\end{tabular}
\caption{Summary of the best-fit parameters and the associated statistics for each spatial template used in our {\it Fermi}--LAT analysis. The final column represents the 95\% upper limit for the extension.}
\label{tab:extent}
\end{table*}
\endgroup

We run extension tests on the best-fit point source utilizing \texttt{GTAnalysis.extension} and the two spatial templates supported in the FermiPy framework, the radial disk and radial Gaussian templates. Both of these extended templates assume a symmetric 2D shape with width parameters radius and sigma, respectively. We allow the position and spectral parameters to vary when finding the best-fit spatial extension in both cases. The best-fit parameters for the extension tests are presented in Table~\ref{tab:extent}. 
The radial Gaussian template is chosen to describe the $\gamma$-ray morphology and is found to have TS$_{\text{ext}}$ = 34.9 (TS$_{\text{ext}} = 2 \times \ln\big(\frac{L_{\text{ext}}}{L_{\text{ps}}}\big)$) and an extension $\sigma = 0.60^{\circ} \pm 0.10^{\circ}$ at the location R.A., Dec. = 286.99\,$^\circ$, +6.38\,$^\circ$ (J2000) with a 95\% uncertainty for the centroid position that is {\bf $r_{95}= 0.30\,^\circ$}. 
The TS and best-fit power-law index for the radial Gaussian source is 37.4 and $\Gamma_\gamma = 1.80 \pm 0.10$, respectively.

The results reported here are in reasonable agreement with the prior {\it Fermi}--LAT analysis from \citet{li2021investigating}, where they found that a radial disk template with $r = 0.51 \pm 0.02 ^\circ$ located at R.A., Dec. = 286.88\,$^\circ$, +6.29\,$^\circ$ (J2000) best characterized the extended $\gamma$-ray emission along with a power-law spectrum best-fit with $\Gamma_\gamma = 1.60 \pm 0.20$. 
The integrated energy flux for the 30\,GeV--2\,TeV energy band is 3.40 $\pm$ 0.90 $\times 10^{-11}$\,erg cm$^{-2}$ s$^{-1}$, consistent with \citet{li2021investigating}.

\section{VERITAS data analysis}\label{sec:veritas_analysis}

VERITAS (the Very Energetic Radiation Imaging Telescope Array System) is an array of four imaging atmospheric Cherenkov telescopes (IACTs) located at the Fred Lawrence Whipple Observatory in Arizona, USA \citep{weekes2002veritas}. 
Each telescope consists of a 12-m diameter reflector and a camera of 499 photomultiplier tube (PMT) pixels, covering a field of view of $3.5^{\circ}$.
VERITAS is sensitive to photons in the energy range from 100 GeV to 50 TeV with an optimal angular resolution of $0.08^{\circ}$ ($68\%$ containment radius) above 1 TeV.
The VERITAS data used in this study are collected during the period between 2009 and 2022.
After data quality selection, an effective total of 128 hours of exposure is available around the location of MGRO J1908+06.

\subsection{VERITAS Background Estimation Methods}

MGRO J1908+06 is an extended source in the TeV band. The associated source 3HWC J1908+063 is reported with a best-fit diffusion radius $\theta_{\mathrm{d}} = 1.78 \pm 0.08 ^{+0.07}_{-0.02}$ degrees \citep{albert2022hawc}.
This is comparable to the $3.5^{\circ}$ diameter of the VERITAS field of view, requiring an alternative to the background-estimation methods that are typically used which estimate the background from the observations by using signal-free regions of the FoV.
Thus, a novel background estimation method (Low-rank Perturbation method, LPM) is developed and described in Section~\ref{sec:low_rank_perturbation_method}. 
The LPM utilizes the distributions of cosmic-ray-like events collected from the MGRO J1908+06 observations that failed the $\gamma$-ray event selection and events from archival $\gamma$-ray-free observations to derive a background estimation for $\gamma$-ray-like background events.
The LPM does not rely on the source morphology assumption and is able to analyze a source with an angular size larger than the field of view of the instrument. An independent background method is used as a cross check for the LPM analysis.
The results of the cross-check analyses can be found in Section \ref{sec:gammapy_ana}.

It is noted that the diffuse $\gamma$-ray emission from the Galactic plane is not included in the background model for the VERITAS analysis, since the Galactic plane emission is not significantly detected in the VERITAS $\gamma$-ray significance map (see Figure \ref{fig:intro_map}) as well as the $\gamma$-ray brightness radial profile (see Figure \ref{fig:radial_profile}).

\begin{figure*}
\gridline{
\fig{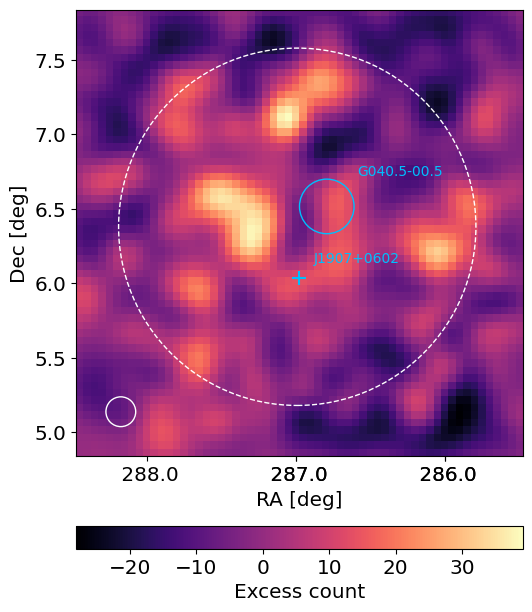}{0.33\textwidth}{(a)}
\fig{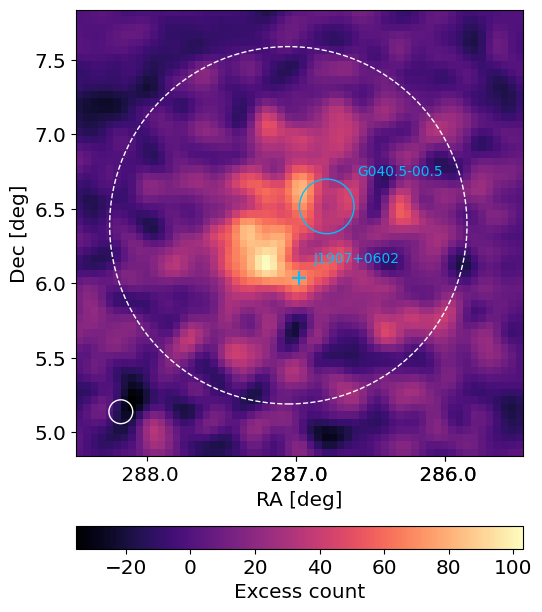}{0.33\textwidth}{(b)}
\fig{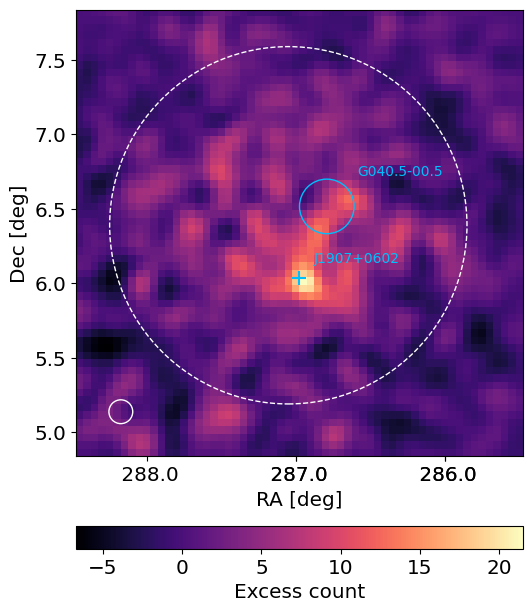}{0.33\textwidth}{(c)}
}
\caption{
(a) {\it Fermi}--LAT $\gamma$-ray residual  map in the energy range $[30,300]$ GeV overlaid with a white dashed circle showing the 95.5\% containment of the $\gamma$-ray photons. 
(b) VERITAS $\gamma$-ray residual map in the energy range $[0.5,2.0]$ TeV overlaid with a white dashed circle showing the extraction region of the $\gamma$-ray spectrum.
(c) VERITAS $\gamma$-ray residual map in the energy range $[2.0,7.9]$ TeV overlaid with a white dashed circle showing the extraction region of the $\gamma$-ray spectrum.
The sky-blue cross indicates the location of PSR J1907+0602, and the sky-blue circle outlines the shell size of SNR G40.5-0.5.
The white solid circles indicate the size of PSF of the corresponding instrument.
The coordinates are in J2000 equatorial degrees.
}
\label{fig:vts_flux_map}
\end{figure*}

\subsection{VERITAS Data Analysis Results}

The VERITAS data and the LPM background method allow the measurement of the entirety of MGRO J1908+06 emission covering a large region with a radius up to $2.5^{\circ}$ centered at the 3HWC J1908+063 centroid.
The radial profile of the MGRO J1908+06 emission is shown in Figure \ref{fig:radial_profile}.
An integrated VERITAS significance map is shown in Figure \ref{fig:intro_map} in the energy range $[0.5,7.9]$ TeV.
Figure \ref{fig:vts_flux_map} (b) and Figure \ref{fig:vts_flux_map} (c) show the energy-dependent $\gamma$-ray residual excess map of MGRO J1908+06 in the VERITAS data for $[0.5,2.0]$, and $[2.0,7.9]$ TeV bins, respectively.
In the $[0.5,2.0]$ TeV bin, the core emission is near but not centered around the location of PSR J1907+0602.
The core emission then becomes more concentrated around the pulsar at the higher energies in $[2.0,7.9]$ TeV.
The energy-dependent morphology of MGRO J1908+06 has also been reported by \cite{kostunin2021revisiting} and showed a similar trend.
After correcting for the effect of off-axis $\gamma$-ray acceptance, a Gaussian model is fit to the $\gamma$-ray morphology in the energy ranges of $[0.5,2.0]$ TeV and $[2.0,7.9]$ TeV.
The energy-dependent Gaussian centroid locations and radii are shown in Table \ref{tab:MW_distance_2_PSR}.

\begingroup
\begin{table}[!hbp]
\scalebox{0.85}{
\hspace{-2.5cm}
\begin{tabular}{ccccc}
\hline
\hline
\ Energy [TeV] & Count & Background & $E^{2} \frac{dN}{dE}\ [\mathrm{TeV}\ \mathrm{cm}^{-2}\ \mathrm{s}^{-1}]$ & $\sigma$ \
\\
\hline
0.50-0.75 & 36435 & $34696 \pm 190 \pm 303$ & $(9.2 \pm 1.1 \pm 2.0)\times 10^{-12}$ & 4.9 \\
0.75-1.26 & 19851 & $18068 \pm 140 \pm 170$ & $(10.1 \pm 0.8 \pm 1.0)\times 10^{-12}$ & 8.1 \\
1.26-2.00 & 9448 & $7814 \pm 97 \pm 187$ & $(12.1 \pm 0.7 \pm 1.3)\times 10^{-12}$ & 7.7 \\
2.00-3.16 & 3803 & $3049 \pm 61 \pm 79$ & $(7.9 \pm 0.6 \pm 0.8)\times 10^{-12}$ & 7.5 \\
3.16-5.01 & 1357 & $984 \pm 36 \pm 55$ & $(5.9 \pm 0.7 \pm 0.9)\times 10^{-12}$ & 5.6 \\
5.01-7.94 & 547 & $290 \pm 23 \pm 20$ & $(6.6 \pm 1.1 \pm 0.5)\times 10^{-12}$ & 8.2 \\
\hline
\hline
\end{tabular}}
\caption{The energy-dependent VERITAS data count, background (with statistical and systematic uncertainties), flux (with statistical and systematic uncertainties), and detection significance $\sigma$ in the region of interest with a radius of $1.2^{\circ}$ centered at the 3HWC J1908+063 centroid.}
\label{tab:VTS_flux_RoI}
\end{table}
\endgroup

\begin{figure}[htbp!]
\centering
\includegraphics[width=1.0\linewidth]{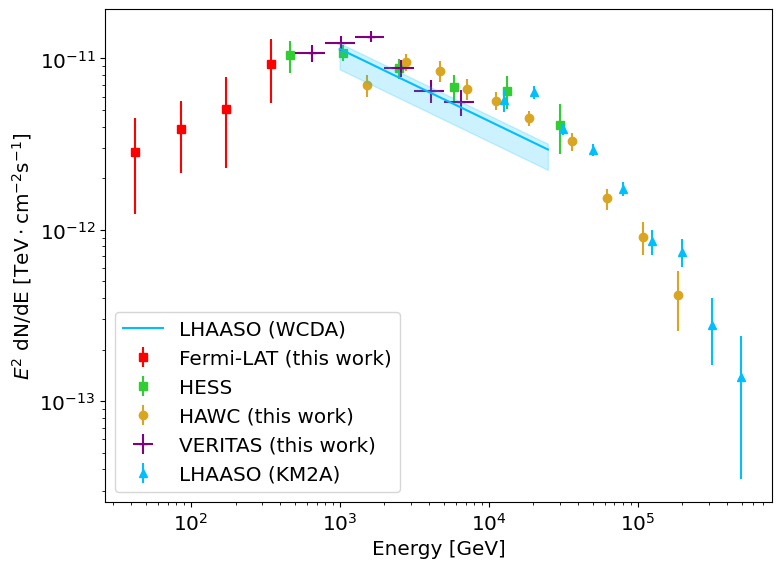}
\caption{
The multiwavelength spectrum of MGRO J1908+06. 
The red points represent the data in this work for {\it Fermi}--LAT data in $[30,300]$ GeV, the purple points represent VERITAS data in $[0.5,7.9]$ TeV, the lime green points represent H.E.S.S. data \citep{kostunin2021revisiting}, the golden points represent HAWC data in $[1.1,246]$ TeV, and the sky blue points represent LHAASO data in $[10,625]$ TeV \citep{cao2021ultrahigh,cao2023first}.
The {\it Fermi}--LAT data are truncated at 300 GeV because of the low statistics at the higher energies.}
\label{fig:MW_spectrum}
\end{figure}

An analysis region centered at the 3HWC J1908+063 centroid R.A., Dec. $= 287.05^{\circ}, 6.39^{\circ}$ \citep{albert2022hawc} with a radius of $1.2^{\circ}$ is used to study the spectrum of MGRO J1908+06.
It is assumed that the $\gamma$-ray morphology reflects the underlying particle distribution, thus, the VERITAS analysis region is designed to cover an area that contains the majority of emissions in {\it Fermi}--LAT and HAWC data. The VERITAS analysis region is highlighted by the white dashed circle in Figure \ref{fig:vts_flux_map} (b) and (c).
The VERITAS flux is calculated from the events included within the region.
The breakdown information in the region of analysis, including data count, background and uncertainties, are summarized in Table \ref{tab:VTS_flux_RoI}.
The VERITAS spectrum of MGRO J1908+06 is shown in Figure \ref{fig:MW_spectrum} along with the spectra of {\it Fermi}--LAT (Section~\ref{sec:fermi_analysis}), HAWC (Section~\ref{sec:hawc_analysis}), and LHAASO data \citep{cao2021ultrahigh,cao2023first}.

\section{HAWC data analysis}\label{sec:hawc_analysis}

The High Altitude Water Cherenkov (HAWC) gamma-ray observatory located in Puebla, Mexico surveys the sky in the 300 GeV to $>$ 100 TeV energy range and has been able to detect the source 3HWC J1908+063 which is $0.29^{\circ}$ apart from MGRO J1908+06 and that has been reported in \cite{albert2022hawc}. For this work, we use the HAWC Data Pass~5 \cite{pass5} that comprises 2321 days of observations collected between November 2014 and October 2021. The data set is divided in bins according to the fraction of photo-multiplier tubes that are triggered in each shower event which are then sub-divided into 12 quarter-decade energy bins covering the 0.316-316 TeV range. This is performed using the ground parameter method presented in \cite{crab-gp}, which uses the fit to the lateral distribution function to measure the charge density 40 meters from the shower core, along with the zenith angle of the air shower, to estimate the energy of the primary gamma ray. 

To perform the spectral and spatial fit, a maximum likelihood technique is used with the Multi-Mission Maximum Likelihood (3ML) framework \citep{3ml} with the HAWC Accelerated Likelihood (HAL) plug-in \citep{hal}. To determine the significance of the fit, we use the likelihood ratio test statistic (TS) which is defined by 
\begin{equation}\label{eq:ts}
    \rm{TS} = 2\ln \left(\frac{\mathcal{L}_{\rm S+B}}{\mathcal{L}_{\rm B}}\right),
\end{equation}
where $\mathcal{L}_{\rm S+B}$ is the maximum likelihood of a signal plus background model, which depends on the spectral and spatial parameters, and $\mathcal{L}_{\rm B}$ is the maximum likelihood of the background-only hypothesis.

The region of interest to model the emission is centered at J2000 R.A.= 287.70$^\circ$ and Dec.= 5.45$^\circ$ with a radius of 3 degrees which includes 3 sources: 3HWC J1908+063 and the east and west lobes of a nearby microquasar, SS433 \citep{abeysekara2018very}. Additionally we included the Galactic diffuse emission (GDE) contribution. A Gaussian model is chosen to describe MGRO J1908+06, rather than the diffusion-based source model used by \cite{albert2022hawc}. The reason for choosing the Gaussian model over the diffusion-based source model is that the characteristic age of the PSR J1907+0602 $(19.5\ \mathrm{kyr})$ indicates that the system is evolved but not old enough for the diffusion process to be the dominant mode of particle transport. 
The spectrum of MGRO J1908+06 is modeled as a log-parabola:
\begin{equation}
    \frac{dN}{dE}=K\left(\frac{E}{10\,\rm{TeV}}\right)^{-\alpha -\beta\ln(E/10\,\rm{TeV})},
\end{equation}
where $K$ is the normalization flux, $\alpha$ is the spectral index and $\beta$ is the curvature; these three parameters are left free in the fit.

The SS433 lobes are modeled as point-like sources with fixed position \citep{ss433} and with power-law spectra:
\begin{equation}
    \frac{dN}{dE} = K\left(\frac{E}{10\,\rm{TeV}} \right)^{-2.0},
\end{equation}
The GDE is modeled with a Gaussian distribution centered in the Galactic plane, and described by  power-law spectrum with index -2.75 \citep{gde1,gde2}; both the Gaussian width $\sigma$ and the normalization flux are left free in the fit.

\subsection{HAWC Data Analysis Results}

\begin{figure*}
\gridline{
\fig{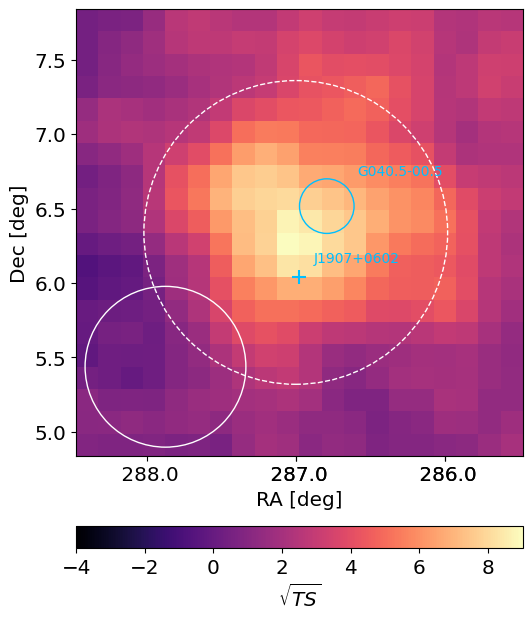}{0.33\textwidth}{(a)}
\fig{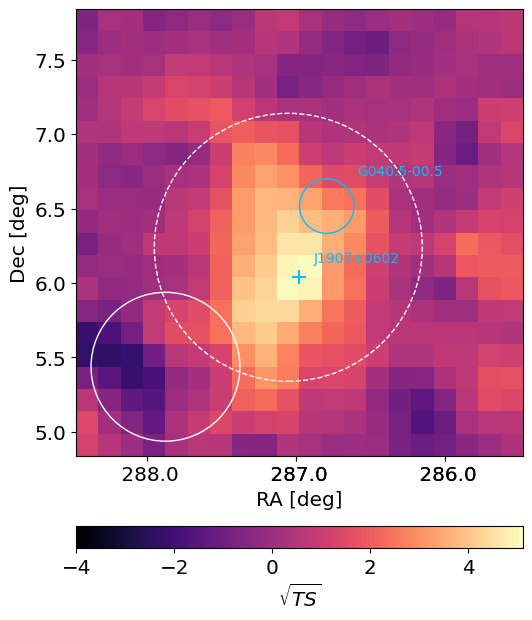}{0.33\textwidth}{(b)}
\fig{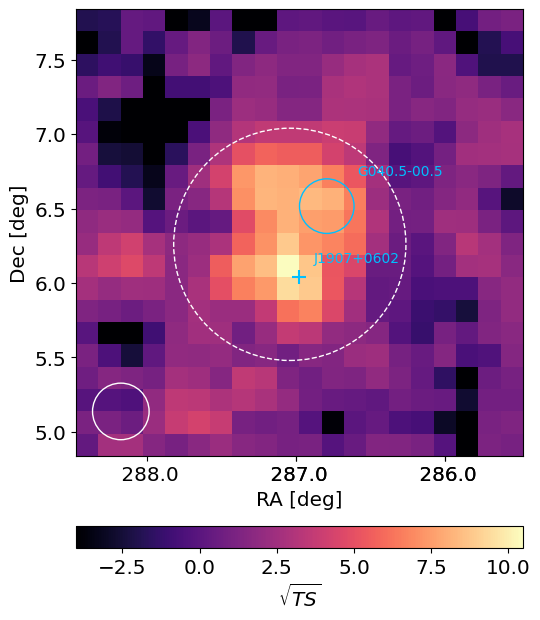}{0.33\textwidth}{(c)}
}
\caption{
HAWC $\gamma$-ray significance  map in the energy range (a) $[3.16,10]$ TeV, (b) $[10,31.6]$ TeV, and (c) $[31.6,100]$ TeV, overlaid with the white a dashed circle showing the 95.5\% containment of the $\gamma$-ray photons in each energy range.
The sky-blue cross indicates the location of PSR J1907+0602, and the sky-blue circle outlines the shell sizes of the SNR G40.5-0.5.
The white solid circles indicate the size of PSF of the HAWC instrument in the corresponding energy range.
The coordinates are in J2000 equatorial degrees.
}
\label{fig:hwc_flux_map}
\end{figure*}

The HAWC $\gamma$-ray significance maps in the energy range $[3.16,100]$ TeV are shown in Figure \ref{fig:hwc_flux_map}.
These HAWC maps show a consistent trend of correlation between the TeV emission and the pulsar PSR J1907+0602, where the $\gamma$-ray emission is seen to be more concentrated around the pulsar at higher energies.
For MGRO J1908+06, the best-fit normalization flux is $K=(4.82\pm 0.19)\times 10^{-14}\;\rm{TeV}^{-1}\,\rm{cm}^{-2}\,\rm{s}^{-1}$, spectral index $\alpha=2.320\pm 0.022$, curvature $\beta = 0.140 \pm 0.015$ and extension $\sigma = 0.^{\circ}475\pm 0.^{\circ}013$. The list of best-fit parameters, including those of the other three sources and systematic uncertainties, is shown in Table \ref{tab:hawc-res}.
The resulting $TS$ values for MGRO J1908+06, SS433E, SS433W and the GDE are 2050.05, 44.0, 63.7 and 276.9, respectively.
In order to obtain the energy-dependent morphology of the source, the Gaussian model is fitted to the HAWC data in five energy bins: $[1.0,3.16,10.0,31.6,100,316]$ TeV, and the centroid and the extension of the source in three of the energy bins are reported in Table \ref{tab:MW_distance_2_PSR}.

Compared to the previous results reported in \cite{albert2022hawc}, a larger TS is reported in this work, with $\Delta TS=120$. 
The flux normalization also varies by $\sim 50\%$ compared to the previously reported value.
The difference between the flux normalization presented by this work and the one presented by \cite{albert2022hawc} is largely due to the choice of the morphological model. 
In \cite{albert2022hawc}, the nominal spatial model is the diffusion-based source model, which has a long
tail and gives a flux normalization that is $\sim 2  \times$ higher than the normalization given by a Gaussian model with the extent fixed to the IACT extent of $0.44^{\circ}$ at $\gamma$-ray energy in $[1,10]$ TeV.
The higher HAWC flux normalization provided by the diffusion model might be overestimated due to the poor angular resolution in the energy range between 1 and 10 TeV (HAWC $\gamma$-ray PSF is $0.64^{\circ}$ at 1 TeV and $0.50^{\circ}$ at 10 TeV) that could lead to an overestimated morphology tail and an overestimated flux.
The HAWC PSF improves at higher energies ($0.19^{\circ}$ at 30 TeV), and the fluxes reported by the diffusion-based source model and the Gaussian model in \cite{albert2022hawc} become more consistent at $\gamma$-ray energy beyond 30 TeV. Additionally, this work utilizes the HAWC Data Pass 5 \cite{pass5}, which incorporates an updated reconstruction algorithm that offers better sensitivity at lower energies compared to the Data Pass 4 used in \cite{albert2022hawc}.

\begin{table}[]
\centering
\begin{tabular}{l c}
\hline
\hline
Parameter  & Best-fit value \\
\\
\hline
\multicolumn{2}{c}{MGRO J1908+06} \\
    \hline
         $\sigma$ & $0.^{\circ}475\pm (0.^{\circ}013)_{\rm stat} (_{-0.014}^{+0.009})_{\rm sys}$ \\
         $K$ & $[4.82\pm(0.19)_{\rm stat}(_{-1.4}^{+0.44})_{\rm sys}]\times 10^{-14}$\\
         $\alpha$ & $2.320\pm (0.022)_{\rm stat}(_{-0.063}^{+0.045})_{\rm sys}$ \\
         $\beta$ & $0.140 \pm (0.015)_{\rm stat}(_{-0.027}^{+0.083})_{\rm sys}$ \\
         \hline
         \multicolumn{2}{c}{SS 433 east and west lobes} \\
         \hline
         $K_{\rm{SS433E}}$ & $[1.00\pm (0.28)_{\rm stat}(_{-0.16}^{+0.18})_{\rm sys}]\times 10^{-16}$ \\
        $K_{\rm{SS433W}}$ & $[3.45\pm (0.59)_{\rm stat} (_{-0.42}^{+0.13})_{\rm sys}]\times 10^{-16}$ \\
        \hline
        \multicolumn{2}{c}{Galactic Diffuse Emission} \\
        \hline
        $\sigma_{\rm GDE}$ & $1.^{\circ}40\pm (0.^{\circ}11)_{\rm stat}(_{-0.009}^{+0.037})_{\rm sys}$ \\
        $K_{\rm GDE}$ & $[5.74\pm (0.33)_{\rm stat}(_{-0.18}^{+0.12})_{\rm sys}]\times 10^{-15}$ \\
\hline
\hline
\end{tabular}
\caption{HAWC best-fit results. The normalization flux values $K$, $K_{\rm{SS433E}}$, $K_{\rm{SS433W}}$ and $K_{\rm{ GDE}}$ have units of $\rm{TeV}^{-1}\,\rm{cm}^{-2}\,\rm{s}^{-1}$.
}
\label{tab:hawc-res}
\end{table}

\section{Emission models for MGRO J1908+06}
\label{sec:mgro_j1908_models}

\subsection{Leptonic PWN model}
\label{sec:leptonic_model}
 
In this section, the TeV emission in $[0.03,625]$ TeV is assumed to be the IC emission from a PWN.
The multiwavelength spectral energy distribution (SED) data in Figure \ref{fig:MW_spectrum} are fitted using the NAIMA python package \citep{naima}.
The electron spectral model is
\begin{equation}
\begin{aligned}
&f(d_{\mathrm{PSR}},E)
\\ 
&=\begin{cases}
A(d_{\mathrm{PSR}})\left(\frac{E}{E_{0}}\right)^{-\alpha}
, & \text{if } E < E_{\mathrm{break}} \\
A(d_{\mathrm{PSR}})\left(\frac{E_{\mathrm{break}}}{E_{0}}\right)^{\beta}
\left(\frac{E}{E_{0}}\right)^{-\alpha-\beta}
, & \text{if } E > E_{\mathrm{break}}
\end{cases}
\end{aligned}
\label{eq:el_spectrum}
\end{equation}
where $E$ is the electron energy, $d_{\mathrm{PSR}}$ is the distance between the pulsar and the observer, $E_{0}=1$ TeV is the reference energy, $A(d_{\mathrm{PSR}})$ is the distance-dependent number of electrons per unit energy, $E_{\mathrm{break}}$ is the synchrotron cooling break, $\alpha=2$ is the electron injection index, and $\beta$ is the spectral softening factor due to radiative cooling.

The far-infrared (FIR) field is expected to have noticeable impact to the IC spectrum in addition to the field of Cosmic Microwave Background (CMB) \citep{breuhaus2021ultra}. 
The data of the target photon field in the far-infrared band (0.27 THz - 8.82 THz) are obtained using the R12 and F98 ISRF models that were published in \cite{porter2017high}.
The energy density spectrum of the target photon field is approximated with a black-body radiation of a temperature of 51.6 K and an energy density of $0.074\ \mathrm{eV}/\mathrm{cm}^{3}$.

\begin{figure*}
\centering
\includegraphics[width=0.8\linewidth]{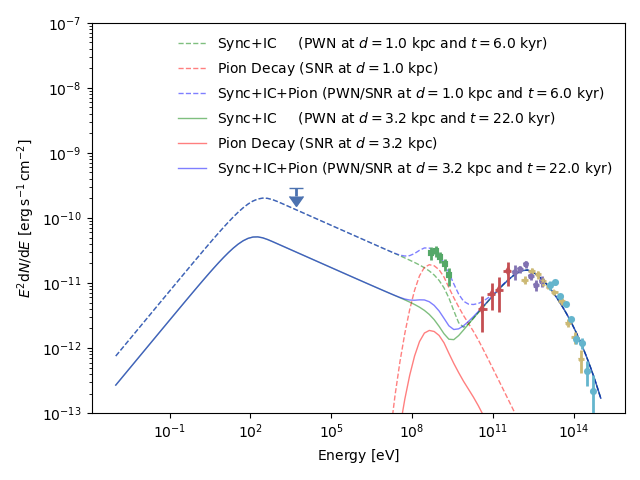}
\caption{
The MW SED data overlaid with the spectra of various emission models.
The blue arrow represents the XMM-Newton upper limit \citep{li2021investigating} scaled by a factor of 2.4 to match the XMM analysis region to the $\gamma$-ray morphology at 1 TeV, the green points represent {\it Fermi}--LAT data in $[0.4,2.8]$ GeV \citep{li2021investigating}, the red points represent {\it Fermi}--LAT data in $[30,300]$ GeV (this work), the purple points represent VERITAS data in $[0.5,7.9]$ TeV (this work), the golden points represent HAWC data in $[1.1,246]$ TeV (this work), and the sky blue points represent LHAASO data in $[10,625]$ TeV \citep{cao2021ultrahigh,cao2023first}.
The green curves show the leptonic emission (synchrotron and inverse Compton) spectra produced by a PWN for different distance scenarios, the red curves show the hadronic emission (pion decay) produced by a SNR with a fixed SN energy $1.0 \times 10^{51}$ erg for different distance scenarios assuming a gas density of $1 \ \mathrm{cm}^{-3}$, and the blue curves represent the combined (synchrotron + inverse Compton + pion decay) spectra.
}
\label{fig:MW_SED_fitting}
\end{figure*}

The best-fit parameters are: $A(d_{\mathrm{PSR}}) = (7.5 \pm 0.8)\times10^{45} \times (d_{\mathrm{PSR}}/\mathrm{kpc})^{2}\ \mathrm{TeV}^{-1}$, $E_{\mathrm{break}} = 17.2 \pm 1.0 \ \mathrm{TeV}$, and $\beta = 1.37 \pm 0.02$.
The result of the fit can be seen in Figure \ref{fig:MW_SED_fitting}.

During the PWN evolution, constant expansion velocity and adiabatic-dominated energy losses are assumed.
In this case, the energy change rate in the PWN is
\begin{equation}
\frac{d E_{\mathrm{PWN}}}{dt}
=
\dot{E}(t) - \frac{E_{\mathrm{PWN}}}{t},
\label{eq:pwn_eq}
\end{equation}
where $E_{\mathrm{PWN}}/t$ is the adiabatic energy loss due to the expansion, and $\dot{E}(t)$ is the time-dependent energy injection from the pulsar spin-down power,
\begin{equation}
    \dot{E}(t) = 
    \dot{E}_{0}
    \left(
    1+\frac{t}{\tau_{0}}
    \right)^{-(n+1)/(n-1)},
\end{equation}
where $n$ is the braking index, $\tau_{0}$ is the spin-down timescale, and $\dot{E}(t=T)=2.8\times 10^{36}$ erg/s is the current spin-down power of PSR J1907+0602, $t=T$ is the true age for PSR J1907+0602.
Using the initial condition that PWN has zero energy at $t=0$, the solution to Equation \ref{eq:pwn_eq} is
\begin{equation}
\epsilon 
=
\frac{(1+x)^{1-y}}{1-y}
-\frac{(1+x)^{2-y}}{x(1-y)(2-y)}
+\frac{1}{x(1-y)(2-y)},
\end{equation}
where $\epsilon=E_{\mathrm{PWN}}/\dot{E}_{0}\tau_{0}$, $x=t/\tau_{0}$, and $y=(n+1)/(n-1)$ \citep{gelfand2009dynamical}.

The electron SED parameter $A(d_{\mathrm{PSR}})$ is connected to the PWN energy $E_{\mathrm{PWN}}$ through the energy fraction in leptons $\chi_{\mathrm{e}}(d_{\mathrm{PSR}},T)$, i.e.
\begin{equation}
    \int_{E_{1}}^{\infty} Ef(d_{\mathrm{PSR}},E)dE = 
    \chi_{\mathrm{e}}(d_{\mathrm{PSR}},T) E_{\mathrm{PWN}}(T), 
\end{equation}
where $E_{1} = 0.1\ \mathrm{GeV}$ is adopted, assuming it to be similar to the inferred value of
the Crab Nebula \citep{busching2008cosmic}, and $f(d_{\mathrm{PSR}},E)$ is the electron spectrum defined in Equation \ref{eq:el_spectrum}.

By equating the true age $T$ and the electron cooling time $t_{\mathrm{cool}}$, the measured $E_{\mathrm{break}}$ in the {\it Fermi}-VERITAS-HAWC multiwavelength data connects the true age $T$ and the energy density of the magnetic field, 
\begin{equation}
T = t_{\mathrm{cool}} \approx
300\left(\frac{E_{\mathrm{break}}}{\mathrm{TeV}}\right)^{-1}
\left(\frac{U_{\mathrm{B}}+U_{\mathrm{rad}}}{\mathrm{eV}\ \mathrm{cm}^{-3}}\right)^{-1}
\ \mathrm{kyr},
\end{equation}
where $U_{\mathrm{rad}}$ is the combined energy density of CMB and FIR fields, and 
\begin{equation}
    U_{\mathrm{B}}(T) = 
    6.24 \times 10^{5} \times
    \frac{1}{8\pi} \times
    \left( \frac{B(T)}{\mu\mathrm{G}} \right)^{2}
    \ \mathrm{eV}\cdot \mathrm{m}^{-3}
\end{equation}
is the magnetic field energy density.
The estimated $U_{\mathrm{B}}(T)$ can be further used to estimate the energy fraction in magnetic field $\chi_{\mathrm{B}}$.
Assuming a uniform magnetic field energy density in a volume 
\begin{equation}
    V_{\mathrm{B}}(d_{\mathrm{PSR}}) = 
    \frac{4\pi}{3} 
    \left(\frac{\pi}{180^{\circ}}\theta_{\mathrm{d}}d_{\mathrm{PSR}}\right)^{3},
\end{equation}
where $\theta_{\mathrm{d}}\sim 0.48^{\circ}$ is the Gaussian radius at $E_{\gamma}=[0.5,7.9]$ TeV, and $d_{\mathrm{PSR}}$ is the distance to the pulsar, one can derive the efficiency $\chi_{\mathrm{B}}(d_{\mathrm{PSR}},T)$ via
\begin{equation} 
    U_{\mathrm{B}}(T) V_{\mathrm{B}}(d_{\mathrm{PSR}}) = \chi_{\mathrm{B}}(d_{\mathrm{PSR}},T) E_{\mathrm{PWN}}(T).
\end{equation}

\begin{figure}
\centering
\includegraphics[width=\linewidth]{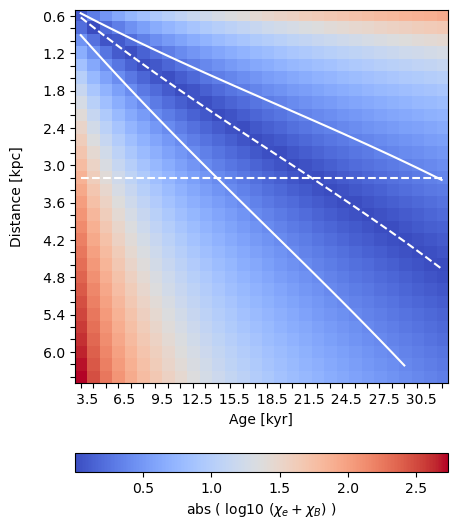}
\caption{
The measured energy fraction sum $\chi_{\mathrm{e}}+\chi_{\mathrm{B}}$ as a function of the pulsar true age and distance to the pulsar, and the distance estimation derived from a dispersion measure \citep{abdo2010psr} is shown as the white horizontal dashed line.
The solid boundaries are derived from the upper and lower limits on the pulsar spin-down time scale, $\tau_{0}\in [0.72,12]$ kyr, and the diagonal dashed curve is derived from a medium value of $\tau_{0}=3$ kyr.
}
\label{fig:test_psr_age_dist}
\end{figure}

Assuming that $\chi_{\mathrm{e}}+\chi_{\mathrm{B}}=1$, one can constrain the pulsar true age and the distance to the pulsar using the observed $\gamma$-ray spectrum and morphology.
Taking the spin-down timescale $\tau_{0}$ to be in the range between an upper limit of 12 kyr \citep{aharonian1995high,abeysekara2017extended} and a lower limit of 0.72 kyr \citep{bucciantini2004effects}, and $n=2.4$ as the apparent braking index averaged from multiple radio pulsars \citep{young2010pulsar}, Figure \ref{fig:test_psr_age_dist} shows the measured energy fraction sum $\chi_{\mathrm{e}}+\chi_{\mathrm{B}}$ as a function of the pulsar true age and distance to the pulsar.
For models of young ages and large distances, the pulsar does not have sufficient energy to support the observed $\gamma$-ray flux and angular extension. 
On the contrary, the models of old ages and small distances could provide too much energy for the observed flux and extension.
If one allows a range of the spin-down timescale $\tau_{0} \in [0.72,12]$ kyr, the two solid lines in Figure \ref{fig:test_psr_age_dist} indicate the parameter space of the most probable age and distance of the pulsar.

The distance to PSR J1907+0602 is estimated to be $3.2 \pm 0.6$ kpc using a dispersion measure \citep{abdo2010psr}. 
Figure \ref{fig:test_psr_age_dist} places an estimate on the true age of the pulsar at $T=22 \pm 9$ kyr and a total energy $W_{\mathrm{e}}=1.5 \times 10^{48}$ erg in electrons above 0.1 GeV. 
Assuming the pulsar true age $T=22$ kyr and a distance $d_{\mathrm{PSR}}=3.2$ kpc, the IC spectral fitting yields $B=5.4 \pm 0.8\ \mu\mathrm{G}$.
Note that the estimated $B$ field is in good agreement with the result presented by \cite{li2021investigating}.
The result of the fit using data from {\it Fermi}--LAT, VERITAS, HAWC, and LHAASO is shown as the solid blue curve in Figure \ref{fig:MW_SED_fitting}.

It should be noted that the null detection of the PWN in X-ray band provides a strong constraint on the age and the distance of the pulsar. 
\cite{li2021investigating} used a region with a radius of $0.34^{\circ}$ around the emission reported by \cite{aharonian2009detection} and obtained an upper limit for MGRO J1908+06 of $1.2 \times 10^{-10} \mathrm{erg} \  \mathrm{s}^{-1} \  \mathrm{cm}^{-2}$.
The $\gamma$-ray morphology of MGRO J1908+06 has an Gaussian radius of $0.53^{\circ}$ at 1 TeV, we thus scale up the XMM upper limit by a factor of $(0.53/0.34)^{2}=2.4$.
The pulsar true age needs to be $>3.5$ kyr, and the distance needs to be $>0.6$ kpc, in order to keep the synchrotron emission below the scaled XMM upper limit (dashed curve in Figure \ref{fig:MW_SED_fitting}).

\subsection{Hadronic SNR model}
\label{sec:hadronic_model}

To investigate the hadronic emission scenario of MGRO J1908+06, the HI and $^{12}$CO (J=1-0) emission line data are studied to estimate the particle density in the region of the $\gamma$-ray source.
The HI data are obtained through the GALFA HI Data Archive of the Arecibo L-band Feed Array (ALFA) at the NAIC Arecibo Observatory 305-meter radio telescope as is presented in \cite{peek2017galfa}.
The $^{12}$CO (J=1-0) data are obtained through the whole-Galaxy CO survey presented in \citet{dame2001milky} and the data files are downloaded from the 1.2 Meter CO Survey Dataverse of Smithsonian Astrophysical Observatory.

\begin{figure*}
\gridline{
\fig{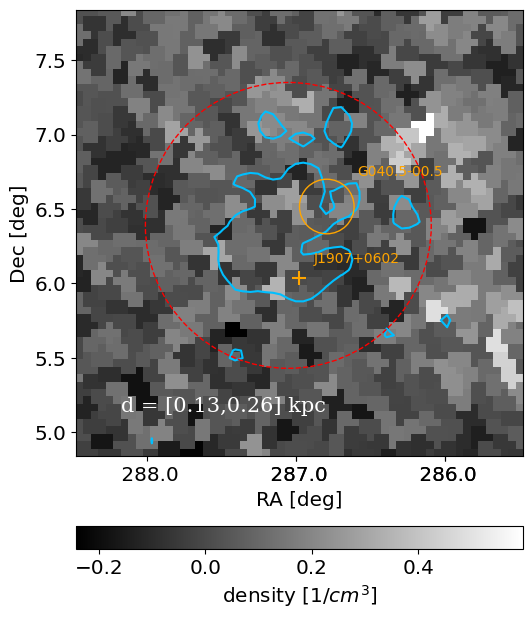}{0.33\textwidth}{(a)}
\fig{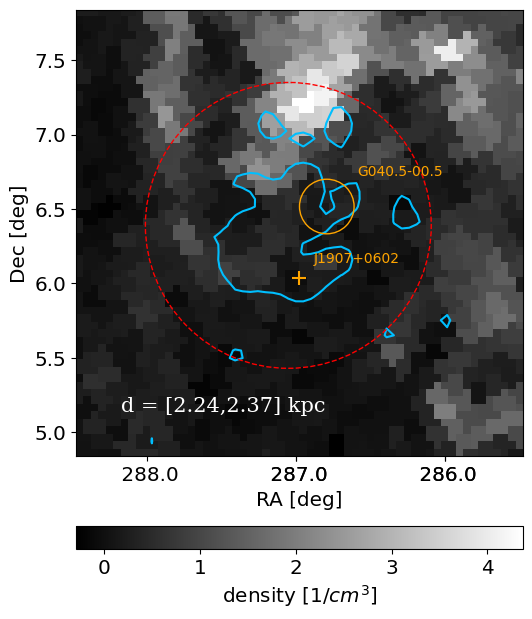}{0.33\textwidth}{(b)}
\fig{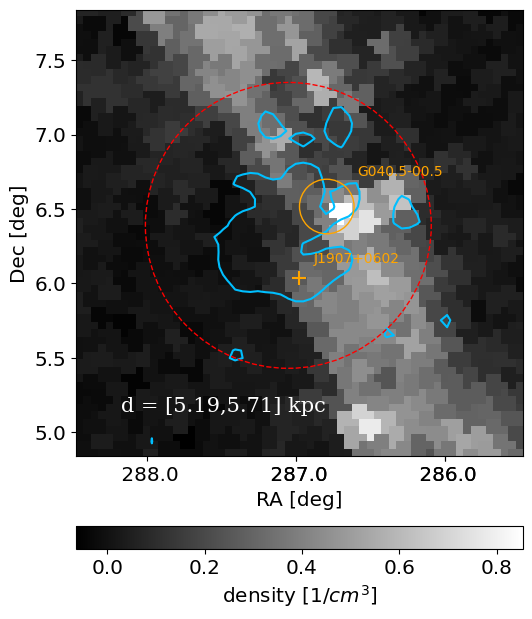}{0.33\textwidth}{(c)}
}
\gridline{
\fig{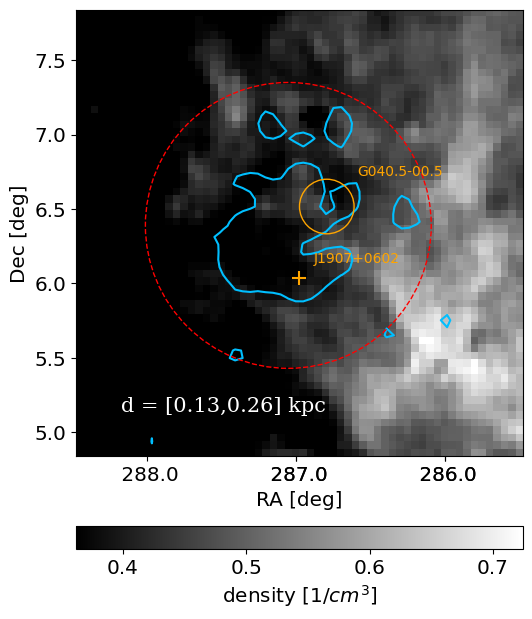}{0.33\textwidth}{(d)}
\fig{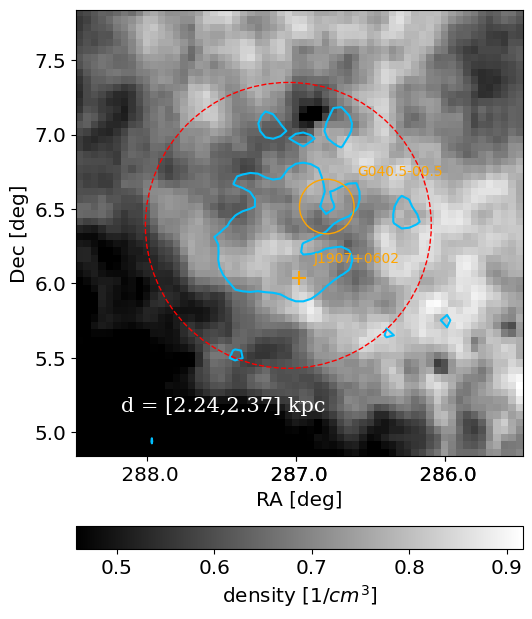}{0.33\textwidth}{(e)}
\fig{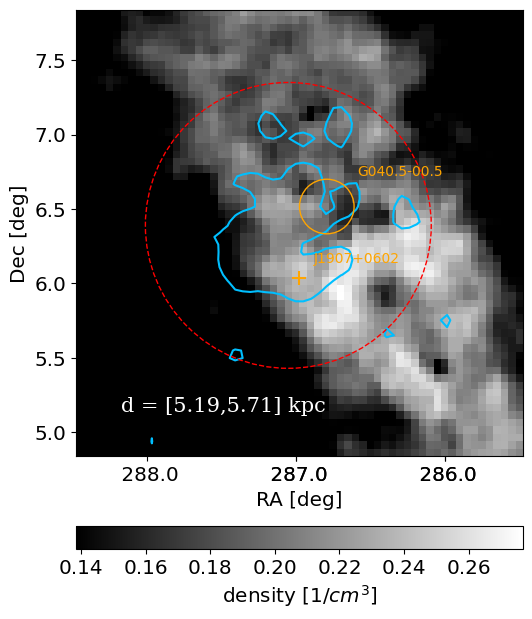}{0.33\textwidth}{(f)}
}
\caption{
The maps of molecular gas density (top row) and atomic gas density (bottom row) in three distance intervals: 0.13-0.26 kpc (a,d), 2.24-2.37 kpc (b,e), and 5.19-5.71 kpc (c,f).
The red dashed circular region is centered at the TeV emission centroid (RA, Dec) = ($287.05^{\circ},6.39^{\circ}$) with a radius of $0.96^{\circ}$ containing $95\%$ of the TeV emission.
The skyblue contour shows the $\gamma$-ray morphology with $>2.5\sigma$ significance in the VERITAS data.
The orange cross indicates the location of PSR J1907+0602, and the orange circle outlines the shell size of the SNR G40.5-0.5.
The coordinates are in J2000 equatorial degrees.
}
\label{fig:gas_density_map}
\end{figure*}

\begin{figure}[htbp!]
\centering
\includegraphics[width=1.0\linewidth]{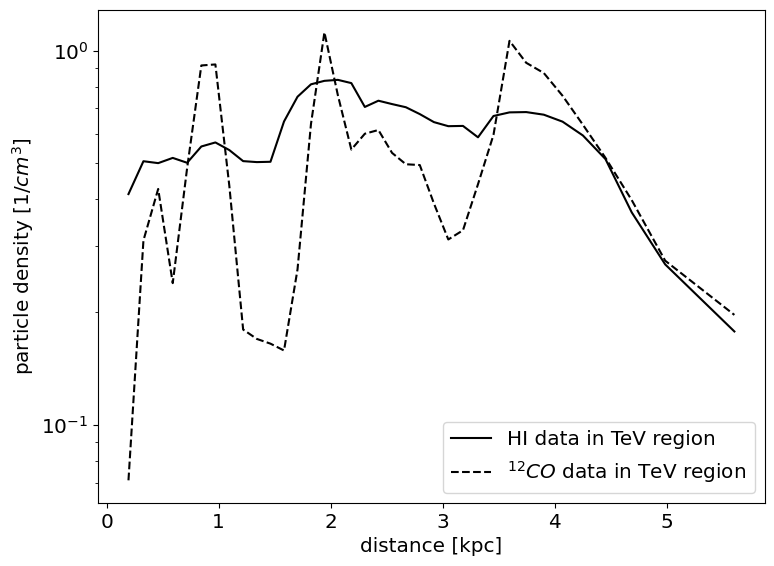}
\caption{
The ambient gas density as a function of distance. 
The dashed curve shows the density estimated using the CO data.
The solid curve shows the density estimated using the HI data.
The densities are measured from the TeV-emission region indicated by the red dashed circle in Figure \ref{fig:gas_density_map}).
}
\label{fig:gas_density_vs_distance}
\end{figure}

The velocity of the atomic and molecular emission is converted to a corresponding distance using the rotation curve model described in \cite{bhattacharjee2014rotation} with the Galacto-centric distance $R_{0}=7.5$ kpc and the rotation velocity of the Sun $V_{0}=190$ km/s.
The particle column density is calculated as $N_{\mathrm{gas}} = X_{\mathrm{gas}} \int T_{\mathrm{gas}} dv\ [\mathrm{cm}^{-2}]$, where $T_{\mathrm{gas}}$ is the brightness temperature of the atomic/molecular gas in K, $v$ is the gas velocity in km/s, $X_{\mathrm{HI}}=1.82 \times 10^{18} \mathrm{cm}^{-2} \ K\  \mathrm{km}\ \mathrm{s}^{-1}$ \citep{duvidovich2020radio}, and $X_{\mathrm{CO}}=2 \times 10^{20} \mathrm{cm}^{-2} \ K\  \mathrm{km}\ \mathrm{s}^{-1}$ \citep{bolatto2013co}.
The integration of the gas emission is performed with velocity intervals of $2$ km/s from $v=2$ km/s to $v=74$ km/s (corresponding to $z=0.13$ kpc and $z=5.71$ kpc), and the particle density is estimated in each velocity interval as $\rho = N/\Delta z$, where $\Delta z$ is the corresponding distance interval.
Three selected maps of atomic and molecular gas densities in the region of MGRO J1908+06 are shown in Figure \ref{fig:gas_density_map}.
The gas density in the region containing the TeV emission is measured. 
The region is centered at the TeV emission centroid (RA,Dec) = ($287.05^{\circ},6.39^{\circ}$) with a radius of $0.96^{\circ}$ containing $95\%$ of the TeV emission and is labeled by the red dashed circle in Figure \ref{fig:gas_density_map}.
The measured gas densities in the region are shown in Figure \ref{fig:gas_density_vs_distance}.
Both the HI data and the CO data show that the gas density in the region is at the level of $\sim 1\ \mathrm{cm}^{-3}$.

To investigate the hadronic emission scenario in which the $\gamma$ rays in $[0.03,625]$ TeV are produced by the $pp$ collision with the ambient gas and the protons are accelerated by SNR G40.5-0.5, the hadronic $\gamma$-ray flux in the TeV regime is modeled by the NAIMA python package \citep{naima} assuming a power-law proton injection spectrum,
\begin{equation}
f_{\mathrm{p}}(E) = C E^{-\gamma} \exp{\left(-E/E_\mathrm{c}\right)}
\end{equation}
where $E$ is the proton energy, $\gamma$ is the power-law index, and $E_\mathrm{c}$ is an exponential cutoff.
The best-fit result for the $\gamma$-ray spectrum at $[0.03,625]$ TeV gives a power-law index of $\gamma = 1.88 \pm 0.03$ and a cutoff energy of $E_\mathrm{c} = 290 \pm 20$ TeV.
The best-fit spectral index of $1.88$ for the proton spectrum is harder than the spectral index of $\sim 2.7$ derived from the $\gamma$-ray observations of W44, Cas A, and IC443 \citep{acciari2009observation,ackermann2013detection}.
This SNR would also need to be located at a distance of $\sim 1.5$ kpc, which is incompatible with the distance estimation (5.5-8.5 kpc) for SNR G40.5-0.5, in order to produce the observed emission flux assuming the SN explosion energy of $\sim 10^{51}$ erg, an energy conversion efficiency of 10\%, and the ambient gas density of $\sim 1\ \mathrm{cm}^{-3}$.
It should be noted that much higher gas density estimations were provided in the previous study by \cite{duvidovich2020radio} by identifying individual clouds at different velocities near SNR G40.5-0.5. 
However, by calculating the $\gamma$-ray flux from the region enclosed by SNR G40.5-0.5 shell, it is found that SNR G40.5-0.5 region only emits $\sim 10\%$ of the total emission from MGRO J1908+06 in the energy range of the VERITAS data and cannot explain the majority of the extended MGRO J1908+06 emission at TeV energies.

In addition to the TeV emission of MGRO J1908+06 at $[0.03,625]$ TeV, the GeV emission at $[0.4,2.8]$ GeV in the region of MGRO J1908+06 (labeled by the green points in Figure \ref{fig:MW_SED_fitting}) discovered by \cite{li2021investigating} is also briefly summarized here.
The best-fit to the GeV emission gives a softer power-law index of $\gamma = 2.81 \pm 0.12$.
The soft GeV $\gamma$-ray spectra from SNRs have been argued to be caused by insufficient turbulence driving ahead of the shock \citep{brose2020cosmic}, resulting in an energy-dependent diffusion coefficient.
\cite{li2021investigating} associated the GeV emission to the possible interaction between cosmic rays and molecular clouds.
Such a scenario is supported by the average gas density of $\sim 45\ \mathrm{cm}^{-3}$ in the GeV emission region at $\sim 8$ kpc \cite{li2021investigating}.

\subsection{A composite SNR/PWN model}
\label{sec:complex_model}

One could consider a scenario in which the GeV emission ($[0.4,2.8]$ GeV) and the TeV emission ($[0.03,625]$ TeV) are produced by the same SNR/PWN system.
In a such scenario, the $[0.03,625]$ TeV emission is produced by the PWN through IC scattering, while the $[0.4,2.8]$ GeV emission is produced by the SNR through $pp$ collisions and subsequent pion decays.
Such a model could exist without violating the scaled XMM upper limit if the SNR/PWN system is located at $\sim 1.0$ kpc with an age of $\sim 6.0$ kyr.
The energy stored in protons above 1 GeV is $W_{\mathrm{p}} = 1.0 \times 10^{50}$ erg for a low gas density of $1\ \mathrm{cm}^{-3}$ in the region.
The energy stored in electrons above 1 GeV is $W_{\mathrm{e}} = 1.5\times 10^{47}$ erg, and the magnetic field $B=10.7 \pm 6.6 \ \mu\mathrm{G}$.
The spectrum of the composite model corresponds to the dashed red curve (hadronic) and the dashed blue curve (hadronic + leptonic) in Figure \ref{fig:MW_SED_fitting}.

\section{Discussion}
\label{sec:discussion}

Based on the observed $\gamma$-ray spectrum and morphology in {\it Fermi}--LAT, VERITAS, and HAWC data, and the estimations provided in Section \ref{sec:mgro_j1908_models}, the possible scenarios for MGRO J1908+06 emission are discussed here.

\begingroup
\begin{table*}[!htb]
\centering
\begin{tabular}{cccccc}
\hline
\hline
\ Instrument & Energy range & R.A. ($^{\circ}$) & Dec. ($^{\circ}$) & $r$ ($^{\circ}$) & $\Delta$ ($^{\circ}$) \
\\
\hline
{\it Fermi}--LAT & 30 GeV - 2 TeV & $286.99 \pm 0.11$ & $6.38 \pm 0.11$ & $0.6 \pm 0.1$ & $0.34 \pm 0.11$ \\
VERITAS & 0.5 TeV - 2.0 TeV & $287.04 \pm 0.01$ & $6.33 \pm 0.01$ & $0.46 \pm 0.01$ & $0.30 \pm 0.01$ \\
VERITAS & 2.0 TeV - 7.9 TeV & $287.00 \pm 0.02$ & $6.27 \pm 0.02$ & $0.42 \pm 0.02$ & $0.23 \pm 0.02$ \\
HAWC & 1 TeV - 3.16 TeV & $287.02 \pm 0.05$ & $6.30 \pm 0.05$ & $0.53 \pm 0.05$ & $0.26 \pm 0.07$ \\
HAWC & 10 TeV - 31.6 TeV & $287.05 \pm 0.03$ & $6.24 \pm 0.03$ & $0.45 \pm 0.02$ & $0.21 \pm 0.04$ \\
HAWC & 100 TeV - 316 TeV & 
$287.02 \pm 0.02$ & $6.20 \pm 0.04$ & $0.40 \pm 0.06$ & $0.16 \pm 0.04$ \\
\hline
PSR J1907+0602 &  -- & 286.98 & 6.04 & -- & -- \\
\hline
\hline
\end{tabular}
\caption{Summary of the best-fit Gaussian centroid and radius $r$, and the angular distance $\Delta$ to PSR J1907+0602 measured by the MW GeV-TeV instruments.}
\label{tab:MW_distance_2_PSR}
\end{table*}
\endgroup

A PWN at early times freely expands until it is crushed by the SNR reverse shock.
The reverse shock disrupts the PWN and often results in an asymmetric morphology of the nebula due to an asymmetric reverse shock, generated by an inhomogeneous medium. The reverse shock compresses the PWN, eventually enabling the pulsar to exit the nebula. 
The separation between the PWN and the pulsar may also be driven by the pulsar's velocity. 
See \cite{blondin2001pulsar, gelfand2009dynamical, sudoh2019tev, giacinti2020halo} for a discussion on the evolution of PWNe.
When the pulsar exits the PWN, the confined PWN electrons continue to lose energy via radiative and adiabatic cooling, forming a relic nebula consisting of low-energy (below the cooling-break) electrons that have a cooling time longer than the age of the system.
A new PWN made up of freshly injected electrons forms near the pulsar which emits high-energy (above cooling-break) photons and have a shorter cooling time than the system age.

A leptonic PWN emission is favored to explain the TeV $\gamma$-ray emission, which is supported by the energy-dependent morphology shown in Figure \ref{fig:vts_flux_map}: the emission with energies above the cooling break in $\gamma$-ray energy ($\sim 1.5$ TeV) is concentrated around the pulsar, while the emission below the cooling-break energy has a larger offset from the pulsar.
The separations between the $\gamma$-ray emission centroid and the location of the pulsar PSR J1907+0602 measured by the MW instruments are summarized in Table \ref{tab:MW_distance_2_PSR}.
The morphology of the $\gamma$-ray emission might be interpreted as a PWN undergoing an interaction with the SNR reverse shock.
The true age of the system is $22 \pm 9$ kyr, and the magnetic field is $5.4\pm 0.8 \mu\mathrm{G}$, assuming the distance to the pulsar is $3.2$ kpc.
The reverse shock interaction and the motion of the pulsar could result in the separation between the relic PWN and the pulsar, creating asymmetric emission below the cooling break ($E_{\gamma}<1.5$ TeV) with an offset from the location of the pulsar.
At higher energies ($E_{\gamma}>1.5$ TeV), the photons are created by high-energy electrons.
These high-energy electrons with short cooling time are freshly produced by the pulsar, which result in more concentrated $\gamma$-ray emission around the pulsar.

The physical distance between the Gaussian centroid of {\it Fermi}--LAT emission reported in Table \ref{tab:extent} and the location of PSR J1907+0602 is 19 pc, if one assumes a distance of 3.2 kpc to the pulsar.
This distance implies a proper motion of $\sim 845$ km/s for the estimated age of 22 kyr, which is quite large compared to the mean 3-D pulsar birth velocity of $400 \pm 265$ km/s measured by a statistical study of 233 pulsar proper motions \citep{hobbs2005statistical}.
However, it should be noted that the asymmetric $\gamma$-ray morphology could be a result of the quenching of the PWN by the reverse shock in addition to the motion of the progenitor star.

Although there is a lack of evidence for the connection between the GeV emission and the TeV emission, a composite system of the PWN of PSR J1907+0602 and the host SNR might explain both the emission in $[0.4,2.8]$ GeV and in $[0.03,625]$ TeV.
This composite system should be $\sim 6$ kyr old and located at $\sim 1$ kpc in order to satisfy the XMM upper limit for the synchrotron emission, maintain a typical SN energy of $\sim 10^{51}$ erg.
The distance estimation for PSR J1907+0602 is $3.2$ kpc and has a nominal error of $20\%$ \citep[$0.6$ kpc,][]{abdo2010psr}. 
The estimated distance $\sim 1$ kpc for the composite system is about four times the standard deviation from the nominal estimation.

\section{Conclusion}
\label{sec:conclusion}

This paper presents the morphology and spectrum of MGRO J1908+06 using multi-wavelength observations across the GeV-TeV band.
The multi-wavelength spectrum and morphology show a clear feature of an energy break at $E_{\gamma} \sim 1.5$ TeV.
The energy break might be due to synchrotron cooling, which separates the cooled and uncooled particle populations.
The multi-wavelength morphology shows that the MGRO J1908+06 emission has an energy dependence, revealing the distribution of particles at different energies.
The {\it Fermi}--LAT and VERITAS data below the energy break reflect a relic PWN of low-energy electrons, while the VERITAS and HAWC data above the energy break reflect a new PWN of high-energy electrons.

The energy-dependent $\gamma$-ray morphology study of the MGRO J1908+06 TeV emission favors a leptonic PWN emission model.
If the PWN is located at $d=3.2 \pm 0.6$ kpc based on the dispersion measure \citep{abdo2010psr}, the best-fit SED gives a true age $T=22\pm 9$ kyr and a magnetic field $B=5.4 \pm 0.8\ \mu\mathrm{G}$.
It is also interesting to note that if the TeV and GeV emission are produced by the same SNR/PWN composite system, such a system could exist at $d \sim 1$ kpc, $T \sim 6$ kyr, and $B\sim 10.7\ \mu\mathrm{G}$.

The synchrotron spectrum predicted from the estimated PWN parameters of PSR J1907+0602 will need new observations from the radio to MeV to verify. Future MeV instruments will be essential for revealing the complete hadronic emission in the MeV-GeV band and the connection between the GeV and TeV emission of MGRO J1908+06.

\newpage

This publication utilizes data from Galactic ALFA HI (GALFA HI) survey data set obtained with the Arecibo L-band Feed Array (ALFA) on the Arecibo 305m telescope. The Arecibo Observatory is operated by SRI International under a cooperative agreement with the National Science Foundation (AST-1100968), and in alliance with Ana G. Méndez-Universidad Metropolitana, and the Universities Space Research Association. The GALFA HI surveys have been funded by the NSF through grants to Columbia University, the University of Wisconsin, and the University of California.

This research is supported by grants from the U.S. Department of Energy Office of Science, the U.S. National Science Foundation and the Smithsonian Institution, by NSERC in Canada, and by the Helmholtz Association in Germany. This research used resources provided by the Open Science Grid, which is supported by the National Science Foundation and the U.S. Department of Energy's Office of Science, and resources of the National Energy Research Scientific Computing Center (NERSC), a U.S. Department of Energy Office of Science User Facility operated under Contract No. DE-AC02-05CH11231.
We acknowledge the excellent work of the technical support staff at the Fred Lawrence Whipple Observatory and at the collaborating institutions in the construction and operation of the instrument. R.S. thanks NSF for support under NSF
grants PHY-1913798 at UCLA and PHY-2110497 at Barnard College. The authors thank J. Gelfand and K. Mori for their constructive comments.

The HAWC collaboration acknowledges the support from: the US National Science Foundation (NSF); the US Department of Energy Office of High-Energy Physics; the Laboratory Directed Research and Development (LDRD) program of Los Alamos National Laboratory; Consejo Nacional de Ciencia y Tecnolog\'{i}a (CONACyT), M\'{e}xico, grants 271051, 232656, 260378, 179588, 254964, 258865, 243290, 132197, A1-S-46288, A1-S-22784, CF-2023-I-645, c\'{a}tedras 873, 1563, 341, 323, Red HAWC, M\'{e}xico; DGAPA-UNAM grants IG101323, IN111716-3, IN111419, IA102019, IN106521, IN110621, IN110521 , IN102223; VIEP-BUAP; PIFI 2012, 2013, PROFOCIE 2014, 2015; the University of Wisconsin Alumni Research Foundation; the Institute of Geophysics, Planetary Physics, and Signatures at Los Alamos National Laboratory; Polish Science Centre grant, DEC-2017/27/B/ST9/02272; Coordinaci\'{o}n de la Investigaci\'{o}n Cient\'{i}fica de la Universidad Michoacana; Royal Society - Newton Advanced Fellowship 180385; Generalitat Valenciana, grant CIDEGENT/2018/034; The Program Management Unit for Human Resources \& Institutional Development, Research and Innovation, NXPO (grant number B16F630069); Coordinaci\'{o}n General Acad\'{e}mica e Innovaci\'{o}n (CGAI-UdeG), PRODEP-SEP UDG-CA-499; Institute of Cosmic Ray Research (ICRR), University of Tokyo. H.F. acknowledges support by NASA under award number 80GSFC21M0002. We also acknowledge the significant contributions over many years of Stefan Westerhoff, Gaurang Yodh and Arnulfo Zepeda Dominguez, all deceased members of the HAWC collaboration. Thanks to Scott Delay, Luciano D\'{i}az and Eduardo Murrieta for technical support.

The \textit{Fermi}--LAT Collaboration acknowledges generous ongoing support
from a number of agencies and institutes that have supported both the
development and the operation of the LAT as well as scientific data analysis.
These include the National Aeronautics and Space Administration and the
Department of Energy in the United States, the Commissariat \`a l'Energie Atomique
and the Centre National de la Recherche Scientifique / Institut National de Physique
Nucl\'eaire et de Physique des Particules in France, the Agenzia Spaziale Italiana
and the Istituto Nazionale di Fisica Nucleare in Italy, the Ministry of Education,
Culture, Sports, Science and Technology (MEXT), High Energy Accelerator Research
Organization (KEK) and Japan Aerospace Exploration Agency (JAXA) in Japan, and
the K.~A.~Wallenberg Foundation, the Swedish Research Council and the
Swedish National Space Board in Sweden. Additional support for science analysis during the operations phase is gratefully
acknowledged from the Istituto Nazionale di Astrofisica in Italy and the Centre
National d'\'Etudes Spatiales in France. This work performed in part under DOE
Contract DE-AC02-76SF00515.


\software{FermiPy \citep[v.1.2.0][]{fermipy2017}, Fermitools: Fermi Science Tools \citep[v2.2.11][]{fermitools2019}, Eventdisplay \citep[v.490][]{maier2017eventdisplay}, NAIMA \citep[v.0.10.0][]{naima}, Gammapy \citep[v.1.0][]{aceroGammapyPythonToolbox2022}}

\appendix

\section{{\it Fermi}--LAT Data Analysis Details}\label{sec:fermi_details}
We perform a binned likelihood analysis with the latest Fermitools and FermiPy Python~3 packages, utilizing the \texttt{P8R3\_SOURCE\_V3} instrument response function (IRF) and account for energy dispersion, to perform data reduction and analysis. We organize the events by PSF type using \texttt{evtype=4,8,16,32} to represent the \texttt{PSF0, PSF1, PSF2}, and \texttt{PSF3} components. A binned likelihood analysis is performed on each event type and then combined into a global likelihood function for the region of interest (ROI) to represent all events\footnote{See FermiPy documentation for details: \url{https://fermipy.readthedocs.io/en/0.6.8/config.html}}. We fit the square 10$\,^\circ$ ROI centered on the PWN position in equatorial coordinates using a pixel bin size $0.05\,^\circ$ and 10 bins per decade in energy (19 total bins). The $\gamma$-ray sky for the ROI is modeled from the latest comprehensive {\it Fermi}--LAT source catalog based on 12\,years of data, 4FGL \citep[data release 3 (DR3),][]{4fgldr3} for point and extended sources\footnote{{\url{https://fermi.gsfc.nasa.gov/ssc/data/access/lat/12yr_catalog/}.}} that are within 15\,$^\circ$ of the ROI center, as well as the latest Galactic diffuse and isotropic diffuse templates (\texttt{gll\_iem\_v07.fits} and \texttt{iso\_P8R3\_SOURCE\_V3\_v1.txt}, respectively)\footnote{LAT background models and appropriate instrument response functions: \url{https://fermi.gsfc.nasa.gov/ssc/data/access/lat/BackgroundModels.html}.}.

With the source model described above, we allow the background components and sources with distances from the ROI center (chosen to be the PSR J1907+0602 position) $\leq3.0$\,$^\circ$ to vary in spectral index and normalization.
The test statistic (TS) value quantifies the significance for a source detection with a given set of location and spectral parameters. The significance of such a detection is proportional to the square root of the TS value \citep[][]{mattox1996}. The TS value is defined to be the natural logarithm of the ratio of the likelihood of one hypothesis (e.g. presence of one additional source) and the likelihood for the null hypothesis (e.g. absence of source):
\begin{equation}
  TS = 2 \times \ln\left({\frac{{\mathcal{L}_{1}}}{{\mathcal{L}_{0}}}}\right),
\end{equation}
TS values $>25$ correspond to a detection significance $> 4 \sigma$ for 4 degrees of freedom (DOF).

\subsection{{\it Fermi}--LAT Systematic Effect Study}
\begin{figure}
\centering
\includegraphics[width=0.6\textwidth]{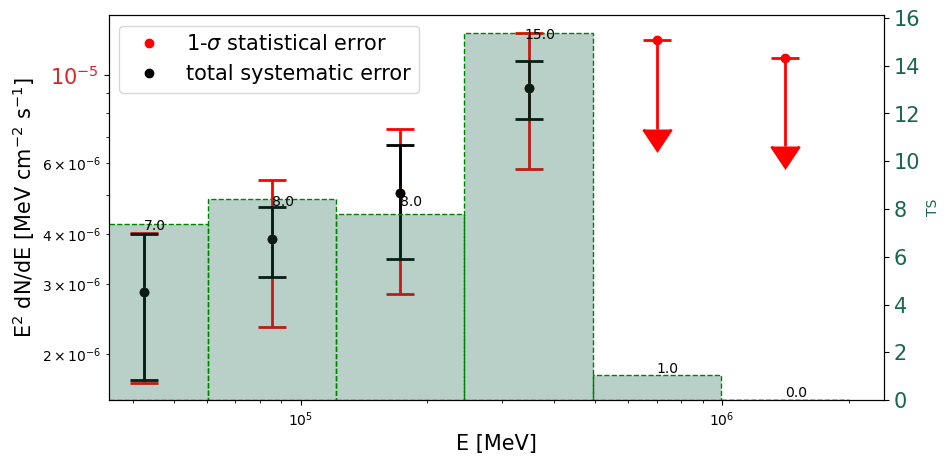}
\caption{The Fermi--LAT SED flux data between 30\,GeV to 2\,TeV including statistical and systematic errors. The upper limits are plotted at the 95\% confidence level. The green histogram displays the TS value for each energy bin.}
\label{fig:sys_errs}
\end{figure}

We account for systematic uncertainties introduced by the choice of the interstellar emission model (IEM) and the IRFs, which mainly affect the spectrum of the measured $\gamma$-ray emission. We have followed the prescription developed in \citet{depalma2013,acero2016}, based on generating eight alternative IEMs using a different approach than the standard IEM \citep[see][for details]{acero2016}. For this analysis, we employ the eight alternative IEMs (aIEMs) that were generated for use on Pass 8 data in the {\it Fermi} Galactic Extended Source Catalog \citep[FGES,][]{ackermann2017}. The extended $\gamma$-ray source coincident with MGRO~J1908+06 is refit with each aIEM to obtain a set of eight values for the spectral flux that we compare to the standard model following equation (5) in \citet{acero2016}. We estimate the systematic uncertainties from the effective area\footnote{\url{https://fermi.gsfc.nasa.gov/ssc/data/analysis/scitools/Aeff_Systematics.html}} while enabling energy dispersion. For energies between 30\,GeV and 100\,GeV, the uncertainty from the effective area is $\pm 3\%$. Beyond 100\,GeV, the uncertainty from the effective area increases as: $\pm (3\% + 12\% \times (\log({\frac{E}{\texttt{MeV}})}-5))$. The IEM and IRF systematic errors are taken to be independent of each other, so we combine the values using the quadratic sum to represent the total systematic error. We find that the total systematic error is comparable to the statistical error, primarily from the increasing uncertainty on the effective area with energy. The $1\,\sigma$ statistical uncertainty remains the largest source of error, however, so we only include statistical errors in this report. 
Figure~\ref{fig:sys_errs} displays both statistical and systematic errors for reference.

\section{VERITAS Data Analysis Details}

\subsection{Low-rank Perturbation Method}
\label{sec:low_rank_perturbation_method}

The Low-rank Perturbation Method (LPM) is the primary analysis method for background prediction used in this paper.
The LPM predicts the background for $\gamma$-ray-like events using the event distribution in the cosmic-ray-like region in the parameter space of mean reduced scaled length ($MSCL$) and mean reduced scaled width ($MSCW$).
The mean reduced scaled length ($MSCL$) is defined as
\begin{equation}
    MSCL = \frac{1}{N} \sum_{i=1}^{N}
    \frac{l_{\mathrm{obs},i}(s,R)-\bar{l}_{\mathrm{sim}}(s,R)}{\sigma_{l,\mathrm{sim}}(s,R)}
\end{equation}
, and the mean reduced scaled width ($MSCW$) is defined as
\begin{equation}
    MSCW = \frac{1}{N} \sum_{i=1}^{N}
    \frac{w_{\mathrm{obs},i}(s,R)-\bar{w}_{\mathrm{sim}}(s,R)}{\sigma_{w,\mathrm{sim}}(s,R)},
\end{equation}
where N is the number of telescopes with images passing selection cuts, $l_{\mathrm{obs},i}$ ($w_{\mathrm{obs},i}$) is the shower length (width) observed by the $i$th telescope, $\bar{l}_{sim}$ ($\bar{w}_{sim}$) is the expected length (width) of the simulated events, and $\sigma_{l,\mathrm{sim}}$ ($\sigma_{w,\mathrm{sim}}$) is the 90 \% containment variation of the length and width of the simulated events.
The length and the width of the shower are binned in shower size $s$ (the summation of the charge of all the pixels of the image) and impact parameter $R$ (the distance from the telescope to the shower axis).

The events passing the mean reduced scaled parameters cuts ($MSCL<0.6$ and $MSCW<0.6$) are the $\gamma$-ray-like events, and the events failing the mean reduced scaled parameters cuts are the cosmic-ray-like events.
The event distribution in the binned $MSCL-MSCW$ space are the "matrix", in which the matrix element $M_{ij}$ is the count of the events in the bin with index $i$ labeling the $MSCL$ dimension and $j$ the $MSCW$ dimension.

An OFF matrix $M^{\mathrm{OFF}}$ is assembled from a set of $\gamma$-ray-free archival observations with observing zenith and azimuth angles matching the ON observation matrix $M^{\mathrm{ON}}$.
The LPM method predicts the difference $\delta M_{ij} = M^{\mathrm{ON}}-M^{\mathrm{OFF}}$ in the $\gamma$-like region by assuming that $M_{ON}$ is a low rank matrix and minimizing the Frobenius norm $\lVert M^{\mathrm{ON}}-M^{\mathrm{OFF}}- \delta M \rVert_F$ in the cosmic-ray-like region.

The procedure of the method is following.
Given an OFF-run matrix $M^{\mathrm{OFF}}$, the Singular Value Decomposition (SVD) is
\begin{equation}
M^{\mathrm{OFF}}
=
\sum_{k} \sigma_{k}^{\mathrm{OFF}}
\vec{u}_{k}^{\mathrm{OFF}} \vec{v}_{k}^{\mathrm{OFF}},
\end{equation}
where $M^{\mathrm{OFF}}$ is the $\gamma$-ray-free matrix constructed from the OFF runs that match the ON runs in observing conditions and is normalized to the cosmic-ray-like region of the $M^{\mathrm{ON}}$, i.e.
\begin{equation}
    \sum_{ij \in \mathrm{CR-like}} M^{\mathrm{ON}}_{ij} = \sum_{ij \in \mathrm{CR-like}} M^{\mathrm{OFF}}_{ij}.
\end{equation}

A linear perturbation theory can be built around $M^{\mathrm{OFF}}$ to the approximation of background matrix $\tilde{M}^{\mathrm{ON}}$ of the ON runs, i.e. 
\begin{equation}
\tilde{M}^{\mathrm{ON}}_{ij} 
= \sum_{k} \sigma_{k}^{\mathrm{ON}}
\vec{u}_{k}^{\mathrm{ON}} \vec{v}_{k}^{\mathrm{ON}}
\approx M^{\mathrm{OFF}}_{ij} + \delta M_{ij}
\label{eq:perturb_M}
\end{equation}
where $\delta M_{ij} = \sum_{k,n=1}^{d} u_{i}^{(k)}v_{j}^{(n)} t_{kn}$ is the matrix perturbation based on the eigenvectors of $M^{\mathrm{OFF}}$, $u_{i}^{(k)}$ ($v_{j}^{(n)}$) is the entry in the eigenvector $\vec{u}_{k}^{\mathrm{OFF}}$ ($\vec{v}_{n}^{\mathrm{OFF}}$), and the coefficient matrix $t_{kn}$ is the solution that needs to be found.

To find $t_{kn}$, one can expand the perturbations around the singular values and the eigenvectors,
\begin{equation}
\sigma_{k}^{\mathrm{ON}} = \sigma_{k}^{\mathrm{OFF}} + \delta \sigma_{k}
\label{eq:perturb_sigma}
\end{equation}
\begin{equation}
\vec{u}_{k}^{\mathrm{ON}} = \vec{u}_{k}^{\mathrm{OFF}} + \delta \vec{u}_{k}
\label{eq:perturb_u}
\end{equation}
\begin{equation}
\vec{v}_{k}^{\mathrm{ON}} = \vec{v}_{k}^{\mathrm{OFF}} + \delta \vec{v}_{k}
\label{eq:perturb_v}
\end{equation}
The perturbations, $\delta \vec{u}_{k}$ and $\delta \vec{v}_{k}$, can be written as the linear combinations of the $M^{\mathrm{OFF}}$ basis, i.e.
\begin{equation}
\delta \vec{u}_{k} = \sum_{n=1}^{d} C_{kn} \vec{u}_{n}^{\mathrm{OFF}}
\end{equation}
\begin{equation}
\delta \vec{v}_{k} = \sum_{n=1}^{d} D_{kn} \vec{v}_{n}^{\mathrm{OFF}}
\end{equation}
, where $d$ is the number of dominant singular values.
Note that the coefficients need to satisfy $C_{kn} = -C_{nk}$ and $D_{kn} = -D_{nk}$ for $U$ and $V$ matrices to be unitary.

Substitute Equation \ref{eq:perturb_sigma} \ref{eq:perturb_u} \ref{eq:perturb_v} with $\sum_{k} \sigma_{k}^{\mathrm{ON}}
\vec{u}_{k}^{\mathrm{ON}} \vec{v}_{k}^{\mathrm{ON}}$ in Equation \ref{eq:perturb_M}, one can find that, to the first-order approximation, the perturbation is 
\begin{equation}
 \delta M_{ij} =
\sum_{k,n=1}^{d} 
u_{i}^{(k)}v_{j}^{(n)}
\underbrace{
\left[
\delta_{kn} \delta \sigma_{k}+
\sigma_{n}C_{kn} - \sigma_{k}D_{kn} 
\right]}_{t_{kn}}
\label{eq:first_order_pertrubation_matrix}
\end{equation}
where $\delta_{kn}$ is Kronecker delta function.

The matrix $t$, which contains the perturbations, can be solved by minimizing the regularized Frobenius norm in the CR-like region,
\begin{equation}
\mathrm{Minimize} \
\left(
\sum_{ij \in \mathrm{CR-like}} \Delta_{ij}^{2} + \beta \sum_{k}^{k=d} t_{kk}^{2}
\right)
\label{eq:reg_frobenius_norm}
\end{equation}
where $\Delta_{ij}=\tilde{M}^{\mathrm{ON}}_{ij}-M^{\mathrm{OFF}}_{ij}- \delta M_{ij} = \Delta M_{ij}- \delta M_{ij}$, $t_{kk}$ are the perturbations on the singular values, and $\beta$ is the regularization parameter to be optimized.
The solution to the minimization above can be found by vectorizing Equation \ref{eq:first_order_pertrubation_matrix}, which rearranges the matrices into single-column vectors, i.e. $t_{kn} \rightarrow t_{y}$ and $\Delta M_{ij} \rightarrow \Delta M_{x}$, and Equation \ref{eq:first_order_pertrubation_matrix} becomes
\begin{equation}
\Delta \vec{M} = A \vec{t},
\label{eq:first_order_pertrubation_vector}
\end{equation}
where the products of $u_{i}^{(k)}v_{j}^{(n)}$ and the regularization constant $\beta$ form the element $A_{xy}$, i.e.
\begin{equation}
A = 
\begin{bmatrix}
u_{0}^{(0)}v_{0}^{(0)} & u_{0}^{(0)}v_{0}^{(1)} & \dots & u_{0}^{(k)}v_{0}^{(k)} & \dots & u_{0}^{(k)}v_{0}^{(n)} & \dots \\
u_{0}^{(0)}v_{1}^{(0)} & u_{0}^{(0)}v_{1}^{(1)} & \dots & u_{0}^{(k)}v_{1}^{(k)} & \dots & u_{0}^{(k)}v_{1}^{(n)} & \dots\\
\vdots & \vdots & & \vdots & & \vdots & \\
u_{i}^{(0)}v_{j}^{(0)} & u_{i}^{(0)}v_{j}^{(1)} & \dots & u_{i}^{(k)}v_{j}^{(k)} & \dots & u_{i}^{(k)}v_{j}^{(n)} & \dots\\
\vdots & \vdots & & \vdots & & \vdots & \\
\beta & 0 & & 0 & & 0 & \\
\vdots & \vdots & & \vdots & & \vdots & \\
0 & 0 & & \beta & & 0 & \\
\vdots & \vdots & & \vdots & & \vdots & \\
\end{bmatrix}
\end{equation}
, and 
\begin{equation}
\Delta \vec{M}^{\top} = 
\begin{bmatrix}
\Delta M_{00} & \Delta M_{01} & \cdots & \Delta M_{ij} & \cdots & 0 & \cdots & 0 & \cdots
\end{bmatrix}
\end{equation}
The least-square solution is found to be
\begin{equation}
\vec{t} = (A^{\top}WA)^{-1}A^{\top} W \Delta \vec{M},
\label{eq:least_square_solution}
\end{equation}
where
\begin{equation}
W_{x_{1}x_{2}}= 
\begin{cases}
\delta_{x_{1}x_{2}} \sigma_{x_{1}},
& \text{if } x_{1,2}\in \mathrm{CR-like}\\
0,          & \text{otherwise}
\end{cases}
\end{equation}
where $\sigma_{x} = 1/\sqrt{M^{\mathrm{ON}}_{x}}$ is the statistical uncertainty of the matrix element. 

The resulting perturbations, $\delta \sigma$, $\delta \vec{u}$ and $\delta \vec{v}$, predict the background normalization in the blinded $\gamma$-like region ($\sum_{ij\in \gamma-\mathrm{like}} M_{ij}$).
The optimization for the regularization parameter $\beta$, and binning of the matrices, as well as the dependence on the initial matrix selection, will be discussed in the future publication of the background method.

\subsection{Validation of Background Estimation of VERITAS Data}
\label{sec:vts_validation}

To validate the background estimation for the extended source in the VERITAS analysis, five independent sets of $\gamma$-ray-free data are selected to mimic the ON data (observations of MGRO J1908+06).
The mimic data runs are selected from extragalactic observations of point-like sources with events from the locations of the point-like sources excluded.
The total exposure of a mimic data set is required to be more than $80\%$ of the exposure of the ON observations.
The telescope pointing coordinate (RA, Dec) of each mimic run is artificially changed to the pointing coordinate of the corresponding ON run to recreate a mimic sky map.
The arrival coordinates of a shower event in the mimic data {\bf are} also changed accordingly,
\begin{equation}
(\text{RA,Dec})_{\text{evt}} = (\text{RA,Dec})_{\text{ON}} + (X,Y)_{\text{derot}},
\end{equation}
where $(X,Y)_{\text{derot}}$ is the de-rotated $(X,Y)$ coordinate of the mimic data event in the telescope camera frame.

The background estimations for the 5 mimic data sets are used as 
an estimation of the systematic uncertainty for the ON-data background.
The systematic uncertainty of the background prediction is assessed by computing the root-mean-square of the relative residuals of the mimic data sets after subtracting the background, i.e.
\begin{equation}
\sigma_{\mathrm{syst}}(x,y) = 
\sqrt{ \frac{1}{5} \sum_{i=1}^{5} \left(\frac{D_{i}(x,y)-B_{i}(x,y)}{D_{i}(x,y)}\right)^{2}},
\label{eq:mimic_syst_err}
\end{equation}
where $D_{i}(x,y)$ ($B_{i}(x,y)$) is the data (background) count in the sky map bin $(x,y)$ in the mimic data set $i$.

\begin{figure}
\gridline{
\fig{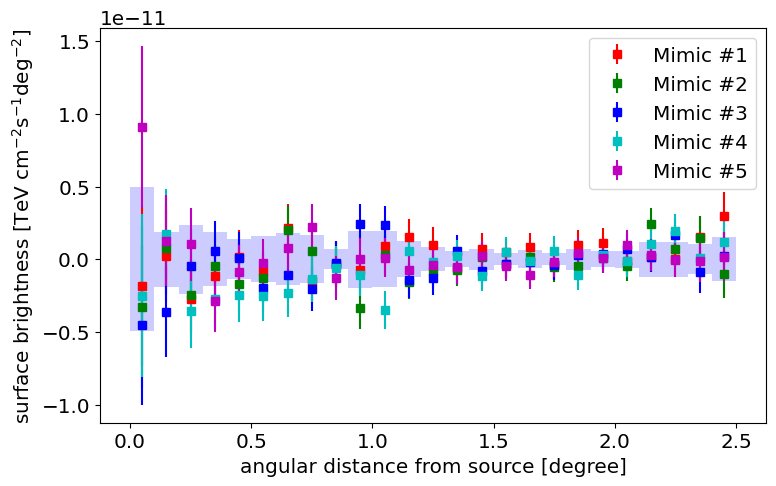}{0.45\textwidth}{(a)}
\fig{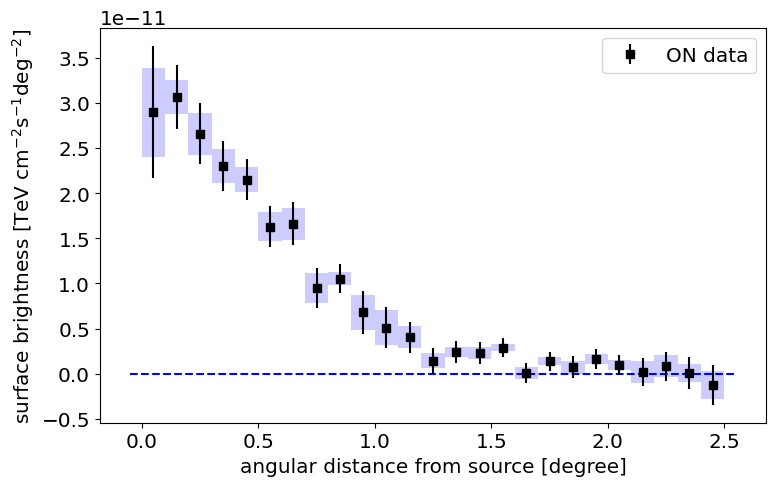}{0.45\textwidth}{(b)}
}
\caption{
The radial profile of the residuals after background subtraction is shown in the unit of surface brightness for (a) the mimic data sets and (b) the ON data set.
The vertical bars represent the data statistical uncertainties, while the blue envelopes are the estimated systematic uncertainties using the root-mean-square of the mimic data profiles.
}
\label{fig:radial_profile}
\end{figure}

Centered at the location of 3HWC J1908+063, Figure \ref{fig:radial_profile} (a) shows the radial profile of the residuals of the mimic data sets after background subtraction (using the background method described in Section \ref{sec:low_rank_perturbation_method}) in the unit of $E^{2}dN/dE$ and the envelope of the residual RMS that represents the systematic uncertainty of background estimation.
Figure \ref{fig:radial_profile} (b) shows the residuals of the ON data set after background subtraction with the envelope of systematic uncertainty derived from the mimic data.

\subsection{Cross-check: Field-of-view (FoV) Background Method}
\label{sec:gammapy_ana}

The FoV Background estimation technique is used as a secondary background estimation method for the VERITAS analysis.
This approach relies on the creation of background models from gamma-ray free data. 
The background models are generated from archival VERITAS data with quality cuts based on realistic analysis data selection cuts for different observing parameters (azimuth, elevation, gamma-hadron separation cut, epoch, night sky background). 

During the background model generation, bright and extended gamma-ray sources are excluded and sources in the VERITAS catalogue are masked with a 0.3 degree radius. 
The background rate is calculated in a (RA, Dec) aligned coordinate system with dependence on energy, radius from the center of the camera, longitude and latitude. 
In this process the full-enclosure (i.e. offset dependent) instrument response functions are interpolated at the average observation parameters. 
In the analysis, observations are stacked onto a map with a bin size of 0.02 degree and a lower energy threshold of 0.8 TeV is used. 
A power-law-based correction is applied to the background rate with the norm and tilt allowed to float freely. Exclusion regions are placed around the HAWC source (1 degree), around the pulsars in the FoV (0.5 degree) and around the two first eastern (e1, e2) and western (w1, w2) jet emission regions of SS 433 (0.5 degree). 

To estimate the systematics and biases in the background prediction, five mimic datasets are created that match the observations in terms of exposure, elevation, azimuth, night-sky-background and epoch as described in Section \ref{sec:vts_validation}.
These files are analyzed with the same pipeline as the observational data. 

\begin{figure}
\centering
\includegraphics[width=0.5\linewidth]{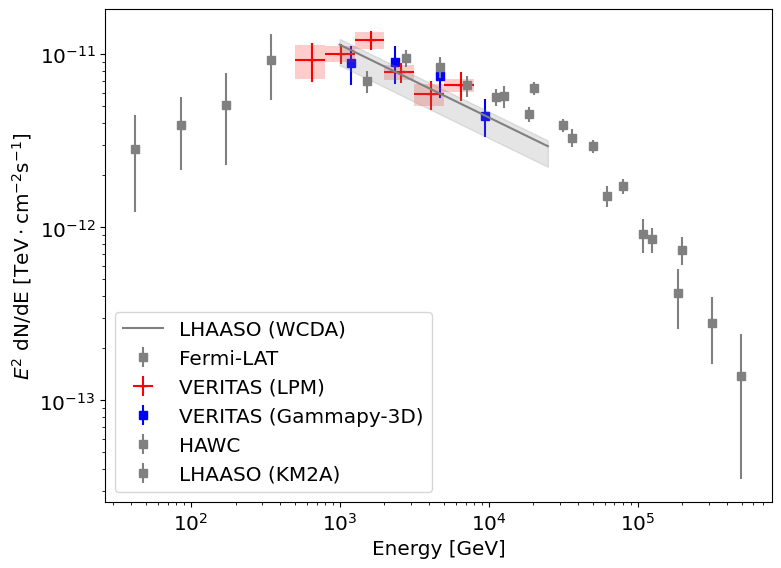}
\caption{
The cross-check VERITAS spectral measurement provided by the Gammapy FoV method (blue points) compared to the spectral measurement provided by the LPM method (red points).
}
\label{fig:VTS_spectrum_crosscheck}
\end{figure}

The comparison between the cross-check provided by the Gammapy FoV method and the nominal spectral measurement provided by the LPM method is shown in Figure \ref{fig:VTS_spectrum_crosscheck}.

\section{HAWC Systematic Uncertainties}
The detector performance and simulations produce a series of systematic uncertainties that are described in detail in \cite{hawc2017} and \cite{hawc2019}. The spectral and spatial parameters with positive and negative shifts are added in quadrature to account for the upward and downward uncertainties, respectively. These uncertainties are included in Table \ref{tab:hawc-res}.

\bibliographystyle{aasjournal}
\bibliography{main}

\end{document}